\documentclass[aps,physrev,preprint,groupedaddress]{revtex4-2}

\usepackage[utf8]{inputenc}
\usepackage{array}
\usepackage{booktabs}
\usepackage{xcolor}
\usepackage{soul}
\usepackage{graphicx}
\usepackage{dcolumn}
\usepackage{bm}
\usepackage{amsmath}
\usepackage{siunitx}

\begin{document}

\title{Characterizing emergent multi-scale dynamics \\in colloidal nanoparticle gels}
\author{William D. Brackett}
\email{brackw@umich.edu}
\affiliation{Department of Chemical Engineering, University of Michigan, Ann Arbor, Michigan, United States \\ McKetta Department of Chemical Engineering, University of Texas at Austin, Austin, Texas, United States
}

\author{Zachary M. Sherman}
\email{zsherm@uw.edu}
\affiliation{Department of Chemical Engineering, University of Washington, Seattle, Washington, United States}

\author{Felix Lehmk\"{u}hler}
\email{felix.lehmkuehler@desy.de}
\affiliation{Deutsches Elektronen-Synchrotron DESY,
22607 Hamburg, Germany}

\author{Thomas M. Truskett}
\email{truskett@umich.edu}
\affiliation{Department of Chemical Engineering, University of Michigan, Ann Arbor, Michigan, United States\\
Biointerfaces Institute, University of Michigan, Ann Arbor, Michigan, United States\\
McKetta Department of Chemical Engineering, University of Texas at Austin, Austin, Texas, United States}

\author{Delia J. Milliron}
\email{milliron@umich.edu}
\affiliation{Department of Chemical Engineering, University of Michigan, Ann Arbor, Michigan, United States\\
Department of Chemistry, University of Michigan, Ann Arbor, Michigan, United States\\
McKetta Department of Chemical Engineering, The University of Texas at Austin, Austin, Texas, United States}

\date{\today}

\begin{abstract}
Colloidal gels assembled from nanoparticles (NPs) are a versatile class of soft network-based materials capable of rich dynamic, mechanical, and even optical or magnetic responses to stimuli. Their behaviors are governed by dynamics of heterogeneous structures coupled across multiple length and time scales. Observable dynamics range from nanoparticle diffusion and clustering to mesoscopic cluster dynamics and interactions to localized or collective network relaxations. Understanding how these hierarchically organized processes relate to macroscopic network properties remains a broad and unresolved problem in soft matter physics. The mechanisms of gel formation can depend sensitively on the pathway and the nature of NP interactions, thus far preventing a unified theoretical bridge between nanoscopic interactions, structural evolution, and network dynamics. Indirect measurement of dynamics using light-scattering techniques provides an experimental means to quantify underlying particle and network motion. The rich dynamic behavior of NP gels warrants consideration of a broad range of models to help interpret nonlinear relaxation phenomena such as anomalous diffusion, nonergodicity, and intrinsically nonequilibrium or mechanically driven dynamics. X-ray photon correlation spectroscopy (XPCS) has emerged as a powerful tool for probing nanoscopic motion in nanoparticle gels, but alone cannot resolve the full spatiotemporal spectrum of dynamics that drive gelation, aging, and network mechanical properties. While in situ rheo-XPCS enables simultaneous probing of nanoscale and bulk mechanical responses, complementary light scattering, microscopy, or simulations can extend spatiotemporal characterization and, consequently, understanding of NP gel network physics.  Implementing a modular model platform with tunable primary nanoparticle features allows systematic variation of nanoscopic characteristics that drive emergent gel responses and inform the development of theoretical models for a wide range of soft, dynamic, nanostructured materials. Gels formed from particles with unique structural proxies, such as electromagnetic coupling in plasmonic NPs, provide additional metrics for model validation and offer opportunities to develop computational methods for the efficient and accurate replication of NP gel properties. The rapid expansion of XPCS capabilities at fourth-generation light sources, combined with complementary tools and robust model systems, positions the field to move beyond descriptive fundamental studies toward the design of nanoparticle gels with adaptive and programmable behaviors.
\end{abstract}

\maketitle

\section{Introduction}
Nanoparticles (NPs) are nanoscopic units of metals, semiconductors, or insulators with characteristic dimensions ranging from 1 to 100 nm. They exhibit behaviors distinct from their bulk material counterparts due to high surface-to-volume ratios, which amplify certain optical, electronic, and catalytic properties. Synthetic tunability of NP size, geometry, surface chemistry, and composition has facilitated nanomaterials-based advances in energy~\cite{pomerantseva2019energy}, optical materials~\cite{li2019stimuli}, catalysis~\cite{kale2014direct}, sensing~\cite{howes2014colloidal}, and therapy~\cite{abadeer2016recent}.

Colloidal NP gels have recently emerged as a versatile class of soft materials. Unlike materials generated via top-down fabrication strategies such as lithographic patterning, NP gels are self-organized, three-dimensional networks that offer unusual combinations of material properties, such as tunable optical response~\cite{brock2006metal,lesnyak20113d,wolf2012quantum,yeom2018chiromagnetic,rosebrock2021spatial,kang2022,kang2023,kang2025}, transport phenomena~\cite{vecchio2022,matter2023,schlenkrich2022revealing}, mechanical properties~\cite{conrad2019increasing,vecchio2022,le2024valence}, and adaptive reconfiguration~\cite{ilg2013stimuli,pastoriza2018plasmonic} through bottom-up, nanoscale design~\cite{matter2020colloidal,sherman2021,green2022}. However, a comprehensive characterization of how nanoscopic features hierarchically translate into gel properties across scales is lacking, which impedes understanding these materials and potentially their development for advanced applications.

Colloidal gelation and aging are governed by dynamics across multiple length and time scales~\cite{zaccarelli2007colloidal,petekidis2021}. Initial NP motion and attractions determine cluster topology~\cite{meakin1988,lazzari2016}. Intra-cluster restructuring and inter-cluster linking drive mesoscopic network growth~\cite{family1985,zaccone2014}. Once arrested, nonequilibrium gels continue to evolve through aging~\cite{jabbari2007gels,guo2011,shahin2012hyper,jain2020,suman2022,chen2023} via microscopic stress redistribution~\cite{ramos2001,bouzid2017}. These processes span several orders of magnitude in space and time~\cite{nixon-luke2020,jungblut2019,jain2022,ofosu2025,song2022}, necessitating integration of multiple complementary experimental approaches. The high X-ray scattering contrast of inorganic NPs~\cite{ahmad2023} enables application of X-ray photon correlation spectroscopy (XPCS) to probe a broad range of gel dynamics. Their tunable surface chemistry also allows the design of model systems with variable dynamic behavior for insight into a broader range of soft materials, including organic and polymeric NP gels relevant to therapeutics, adhesives, biomimetic materials, and more.

In this perspective, we discuss how gaps in the nanoscopic analytical frameworks together with experimental and computational limitations have restricted our ability to fully understand the complex emergent dynamics in colloidal NP gels. We highlight the strengths of XPCS for probing nanoscopic dynamics and how visible-light scattering, differential dynamic microscopy (DDM), particle tracking, network rheology, and simulations can enhance comprehensive analysis across spatiotemporal scales. We further propose modular, molecular-linker-mediated gels as model systems for tuning dynamic properties and for systematically investigating how primary-particle design affects emergent dynamics. Finally, we examine how advances in coherent light sources, ultra-resolution nanoscopy, and improved analysis of nonequilibrium XPCS and rheology, when paired with XPCS (rheo-XPCS), are poised to overcome major challenges in understanding NP gel dynamics.

\section{Background of gel dynamics and open questions }
\begin{figure*}
    \centering
    \includegraphics[width = \textwidth]{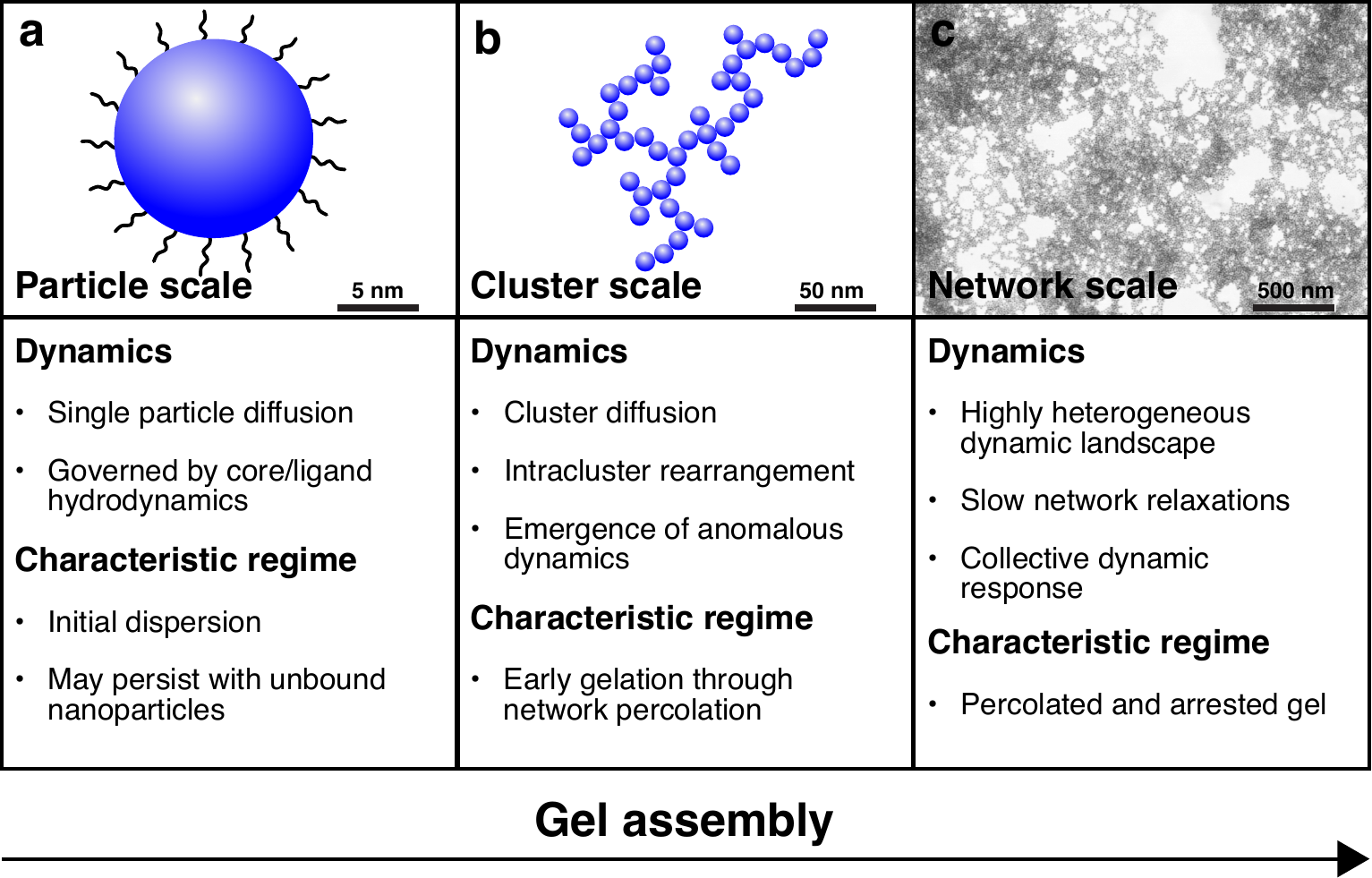}
    \caption{Emergent dynamics and structural scales during NP gelation from (a) dispersion to (b) clusters to (c) networks. Adapted with permission from~\cite{dominguez2020}. Copyright 2020 American Chemical Society.}
    \label{gel_scales}
\end{figure*}

NP gel assembly involves various scale-dependent dynamic processes. Isolated NPs diffuse through the solvent (Figure~\ref{gel_scales}a). Strong interparticle attractions induce structural coarsening of primary NPs into finite clusters with anomalous internal dynamics and attenuated translational mobility (Figure~\ref{gel_scales}b). Growth and merging of clusters due to intercluster attractions lead to percolated networks, which can exhibit significant dynamical and structural heterogeneity and collective multi-scale relaxations underlying bulk mechanical properties of macroscopic gels (Figure~\ref{gel_scales}c).

To achieve the end goal of bottom-up tunability of bulk gel properties and temporally programmable gelation, it is necessary to consider dynamics across the dominant structural scales that emerge during gelation. In this section, we discuss these scale-dependent dynamic processes.

\subsection{Dynamics of nanoparticle building blocks}
The dynamic behavior of dilute NP building blocks represents one of the most fundamental types of motion that can be observed in NP gels. NP geometry (shape and size), ligand shells, and solvent interactions can significantly alter NP dynamics. Understanding motion of individual particles is the first step to developing a broader framework for characterizing gel networks.

Most NPs are colloidally stabilized by solvent interactions with surface-bound ligands and Coulombic repulsions between surface charges. NP surfaces can accommodate hundreds to thousands of ligands, depending on NP size and shape~\cite{villarreal2017}, which influence the local hydrodynamic environment. While dilute dispersions have been studied to probe the dynamic effects of NP surface modifications, hydrodynamic and electrostatic interactions are long-ranged, and within heterogeneous colloidal media such as NP gels, these interactions remain poorly understood. Charge, ligand structure, and solvent interactions can all influence NP mobility and diffusion properties~\cite{moncure2023,hatzis2025, giorgi2019,cui2021modulation} and reinforce the importance of seemingly small differences between primary NPs.

Ligand shells can introduce hydrodynamic drag, solvent layering, and dynamic ligand rearrangements that lead to anomalous dynamics in locally concentrated environments. For instance, diffusion of PEGylated gold NPs through porous hydrogels shows a complex dependence on grafting density~\cite{moncure2023}. High ligand density results in hard-sphere diffusion, while low ligand density results in faster diffusion facilitated by ligand chain relaxations. Charged NP surfaces are surrounded by ion clouds and electric double layers that can increase local solvent friction~\cite{weiss2018} and modulate diffusive behavior through electrostatic interparticle interactions~\cite{giorgi2019}. Ligand shells and surface charge can interact to create a dynamic environment near NP surfaces that depends on solvent properties, local ion concentrations, and ligand-ligand interactions~\cite{hatzis2025,huang2024}. 

In the scope of NP gel networks, initial intuition may be that individual NP dynamics are relatively simple. However, in light of the nanoscale colloidal effects described above, a careful, system-dependent analysis of NP dynamics is demanded. 

\subsection{Cluster formation and heterogeneous dynamics}

As clusters form due to interparticle attractions on a pathway to gelation, colloidal dynamics become \textit{subdiffusive}. Computer simulations, microscopy, and microrheology studies on gels of micron-scale colloids show how this dynamic trend reflects the emergence of dynamically heterogeneous environments~\cite{puertas_dynamical_2004,dibble_structure_2006,dibble_structural_2008,lee2008response,patrick2008direct,tateno2022microscopic}. Highly coordinated or ``caged'' colloids, buried within the interiors of clusters, have coherent dynamics with low mobility. Colloids on the cluster surfaces, in the gaps between clusters, or at network junctions are more mobile. Coarse-grained models of colloidal gels have been simulated to explore how caging and heterogeneous dynamics relate to the mechanism of gel coarsening~\cite{zia2014micro}. Similar models have also predicted qualitative differences in how caging affects the evolution of the shear modulus during gel versus glass formation~\cite{wang2024distinct}, providing physical insights into what distinguishes these two prominent types of structural arrest in disordered materials.

Based on infrared absorption spectroscopy and small-angle X-ray scattering (SAXS) experiments and simulations of plasmonic NPs assembled into linked gel networks by ligand-metal coordination bonds, it was argued that subdiffusive dynamics may be mediated by cluster-surface ``crawling'' in which two bonded NPs swap linked ligands to allow relative particle motion while remaining in contact~\cite{kang2025}. 
Similar surface-bound crawling has been observed for DNA-coated colloids on planar substrates decorated with complementary strands~\cite{zheng2024} and for small gold NPs hydrogen-bonded to larger silica particle surfaces~\cite{cai2022}. 

Although direct visualization of real-space caging and crawling particle motions is not possible for nanoscopic colloids using microscopy, detailed characterization of intermediate-scale dynamics preceding percolation, combined with simulation, offers a route to uncovering how NP and cluster properties influence subsequent gel network formation, emergent scale heterogeneity, and collective mechanical response of NP gels.

\subsection{Arrest pathways and emergent dynamics}
At sufficiently high volume fractions, hard-particle colloidal dispersions that avoid crystallization undergo a dynamic transition from an ergodic fluid to a ``repulsive'' glass as colloids become localized by the excluded volume cage of their nearest neighbors~\cite{bengtzelius1984dynamics,pusey1986phase}. If strong, short-range attractions are also present, an ``attractive'' glass with different interparticle structuring forms as particles instead become localized due to constraints of their attractive bonds with neighboring colloids~\cite{dawson2000higher,zaccarelli2009}. As described below, colloidal gelation, the dominant arrest process that occurs in colloids at low-to-intermediate volume fraction when attractions are sufficiently strong, is comparatively complex and exhibits a strong dependence on pathway~\cite{helgeson2014homogeneous,gao2015microdynamics}. 

Attractive NP gels form fractal, percolated networks that restructure as they evolve toward long-lived arrested states~\cite{cao2010structural,aime2018power,dominguez2020,kang2022,kang2023,kang2025,morozova2023colloidal}. 
Their fractal topology emerges from sequential aggregation of NPs and NP clusters into volume-spanning NP networks. Homogeneous fractals, by definition, possess scale-invariant structural symmetry and statistical properties determined by the route of assembly. The \textit{fractal dimension}, $d_f$, quantifies this architecture by describing the dimensionality of how mass is distributed in NP networks. Such colloidal networks are characterized by $1<d_f<3$, where values nearer to 1 correspond to string-like morphologies and values approaching 3 represent dense coalesced structures~\cite{lazzari2016}.

In dilute NP gels, aggregation (irreversible) or agglomeration (reversible) are often described by two limiting models: diffusion-limited cluster aggregation (DLCA), where NPs stick upon contact, and reaction-limited cluster aggregation (RLCA), where many collisions precede bonding and assembly proceeds more slowly. DLCA kinetics are governed by particle diffusion rates~\cite{jungblut2019,weitz1984}, and percolated fractal gels formed by DLCA pathways tend to be open and porous ($d_f \approx1.8)$. Conversely, RLCA gelation yields structurally denser networks ($d_f\approx2.1$)~\cite{lin1989,weitz1985}. Once clusters reach a critical size, they percolate through the sample volume. The characteristics of early aggregates or agglomerates can significantly affect the properties of percolated networks after initial formation~\cite{zaccone2009elasticity,valadez2013dynamical,zaccone2014,whitaker2019,zhang2019correlated,nabizadeh_network_2024,tsurusawa2019direct}.

The pathway from clusters to percolated gel networks depends on how the gel state is prepared (i.e., protocol) and 
the nature of the interactions between colloids~\cite{zaccarelli2007colloidal,petekidis2021}. If particles are frictionless and have isotropic (centrosymmetric) short-range attractions, they tend to form heterogeneous gels after a quench to a low-temperature (i.e., strong-bonding) state. The heterogeneity emerges due to dynamic arrest of the phase separation which would otherwise drive the system to form macroscopic dilute and colloid-rich fluid phases. Depletion gels represent a common experimental realization of percolated networks that form in this manner~\cite{lu2008,zaccarelli2008gelation}. Quenches to conditions triggering spinodal decomposition by an increase in depletant concentration---i.e., an increased attraction between colloids---initiate a phase separation process that is preempted when the material becomes kinetically trapped in a percolated network of dense, glassy clusters. The heterogeneous structure of such gels often continues to evolve, albeit at much slower rates, after gel formation as the networks continue to coarsen or age. 

In contrast, noncentral interactions between colloids due to surface heterogeneity, roughness, or adhesive contacts~\cite{palangetic2016near,pantina2005elasticity,furst2007yielding,eberle2011dynamical,kim2013gel,valadez2013dynamical} can lead to more rigid bond angles between neighboring colloids and homogeneous gel formation without phase separation at intermediate colloid volume fractions. In such systems, e.g., octadecyl silica NPs with thermoreversible adhesive attractions being a prototypical example, gelation coincides with a rigidity percolation transition which can occur in the stable, homogeneous fluid at conditions far from the spinodal~\cite{eberle2011dynamical}.
Particles with discrete directional attractive sites (``patchy particles'') can form equilibrium homogeneous gels for valence-limited colloids and heterogeneous, arrested-phase-separation gels when particle valence is high~\cite{sciortino_reversible_2011}. Tunability of the gel phase diagram through valence~\cite{bianchi2006phase}, or statistical valence controlled via concentration of linking molecules~\cite{helgeson2014homogeneous,lindquist2016formation,kwon_dynamics_2022}, offers a potentially rich description of gelation but has not yet been extensively experimentally explored.

NP gels also often exhibit nonconformity to a single clustering regime, particularly during aging after percolation. Crossover between RLCA and DLCA is highly dependent on the physical properties and interactions of NPs and early clusters~\cite{meakin1988,kolb1984,sandkuhler2005,family1985}. Experimental work has shown time-salt superposition of the structural crossover from RLCA to DLCA in gelation of charge-stabilized spherical polymer NPs~\cite{wu2013}. This result qualitatively agrees with recent studies demonstrating time-salt superposition of structure and dynamics in covalently-linked gels~\cite{ofosu2025}. However, simulations predict that aggregation of initially reaction-limited patchy anisotropic particles can continue to resemble RLCA well into the percolated regime before exhibiting diffusive limitation~\cite{corezzi2010}. Universal kinetic models have successfully described the gelation behavior of attractive micron-scale colloids~\cite{rouwhorst2020, rouwhorst2020a}, though the applicability of such models to the diverse variety of interactions relevant to NP gelation remains an open question. 

Indeed, developing experimental approaches that allow measurement of the multiscale structural dynamics that accompany network formation~\cite{cho2020, schulz2024a} is essential for understanding and extending existing theoretical frameworks to describe NP gelation.

\subsection{Network structure, mechanical response, aging, and failure}
The linear elastic moduli of {\emph{homogeneous}} fractal colloidal networks, such as octadecyl-grafted
silica NPs in decalin~\cite{guo2010connecting} or tetradecane~\cite{eberle2011dynamical,gordon2017rheology,suman2022} or aqueous solutions of laponite NPs~\cite{chen2020}, 
can be estimated from their localization length using normal-mode analysis~\cite{krall1998,hsiao2014} or mode-coupling theory~\cite{shah2003viscoelasticity,chen2004microscopic,schweizer2007collisions}. 
Effects of heterogeneity and clustering, e.g., which in micron-scale colloidal depletion gels can be directly measured using microscopy~\cite{manley2005glasslike,dibble_structure_2006,lu2008,whitaker2019,royall2021real}, are essential to understanding gel elasticity~\cite{zaccone2009elasticity,whitaker2019} and have also been incorporated into normal-mode analysis and a spring-dashpot model to rationalize sensitivities of the linear viscoelastic response to colloid volume fraction and strength of attractive interactions~\cite{rocklin2021}. These pronounced variations in rheological properties due to heterogeneity cannot be readily explained by trends in local structural metrics such as the average coordination number~\cite{jamali2019multiscale, mangal2024}, which show only modest changes. In thermoreversible gels, it has been experimentally demonstrated that the degree of heterogeneity, and hence the type of rheological aging, can be controlled by quench depth~\cite{suman2022}. Analogously, the structural properties of linker-mediated nanocrystal gels based on competitive metal–ligand equilibria are sensitive to both bond strength and lability~\cite{kang2025}. A comprehensive understanding of these and related effects will require experimental tools capable of characterizing NP network structure and dynamics across scales.

One distinct property of colloidal NP gels that obscures the interpretation of network mechanics from fundamental principles is their tendency to heterogeneously distribute stress across scales. Song et al. show that simple linear tensile strain of a gel network of iron oxide NP gels with four-arm star polymer linkages shifts quiescent collective relaxation of internal stresses to much broader heterogeneous dynamic behavior~\cite{song2022}. Microscopic dynamics have also been shown to precede macroscopic failure. Aime et al.~\cite{Aime2018} postulate that non-affine relaxations in silica NP gels govern network failure under shear. Using static and dynamic light-scattering in an apparatus with stress-controlled rheology, the authors find the gel exhibits a burst of microscopic plastic rearrangements well before gel failure. At small length scales, NP dynamics are plastic and highly spatially heterogeneous despite no signatures of mechanical failure in the macroscopic network. 

In simulations of gels with reversible bonds, continuous reconfiguration causes network-wide coarsening behavior during shear-induced yielding (or even gravitational sedimentation forces), which compact NPs and create osmotic pressure gradients that accelerate gel aging~\cite{johnson2021}. Recent simulation work shows that microscopic colloidal organization predicts which mesoscale particle strands will break upon yielding, and that strand breaking occurs homogeneously in space, consistent with a ductile response~\cite{bhaumik2025yielding}. 

Despite progress in quantifying their hierarchical structures~\cite{smith2024topological}, developing generalizable models for heterogeneous multi-scale gels remains an open challenge~\cite{song2023}. One strategy is to explicitly encode fractal heterogeneity across length scales by introducing hierarchical fractal building blocks with distinct elastic contributions~\cite{bouthier2023}. Alternatively, the microscopic physics of simulated gels can be connected to the hierarchical organization of mechanical elements in a recursive rheological ladder~\cite{Bantawa2023}. Regardless of the approach, experimental access to dynamical properties across multiple spatiotemporal scales is necessary to validate new theoretical models.

\subsection{Model NP gel formers}
Model NP gel-forming systems with different mechanisms for achieving tunable NP attractions have been introduced to investigate gel properties.

Thermally reversible interactions enable exploration of gelation temperature and quench rate. For example, data from experimental studies of polymer-coated silica~\cite{guo2011, bahadur2019, suman2022} or gold~\cite{jain2020, jain2022, schulz2024a}. NPs have helped shape our understanding of emergent gel dynamics during thermal quenches.

Destabilization of charged NPs by adding salt or altering pH leverages tunable solvent environments to control attractions. Commonly studied building blocks in this class of gel-formers include Laponite~\cite{mohammed2022, saito2025,jabbari-farouji2012, jabbari-farouji2012}, silica~\cite{simonsson2018,beelen1994,sogaard2021}, or polymer~\cite{wu2013} NPs. Mixtures of oppositely charged polymer NPs with varying stiffness also exhibit rich gelation behavior~\cite{morozova2023colloidal}.

Though most commonly employed to induce gelation of micron-scale colloids, polymer-mediated depletion interactions have also been used to drive gelation of NPs, e.g. silica NPs~\cite{kim2015depletion} or metal-oxide nanocrystals~\cite{saezcabezas2018,saez2020universal}.  
Limits of the classical depletion theories for describing polymer-induced colloidal attractions between NPs have recently been experimentally probed via small-angle X-ray scattering measurements~\cite{ofosu2025assessing}. One key advantage of depletion interactions for studying NP gelation is the tunability of the depletion attraction via depletant size and concentration and its applicability to NPs of different shapes or compositions~\cite{saez2020universal}.

NPs capped with functionalized ligands can be assembled into gel networks via linkers using dynamic chemistries, e.g., metal-ligand coordination~\cite{lesnyak2010cdte,hitihami2014bottom,singh2015linking,sayevich20163d,song2020programmable,hewa2021electrochemical,wang2021colloidal,kang2022,song2022,kang2023,kang2023a,kang2025,kang2025} or dynamic covalent bonding~\cite{dominguez2020,sherman2023,ofosu2025}. As with depletion gels, the use of linkers to mediate NP assembly separates, at least in part, the properties of the colloidal building blocks from the primary driver force for gelation, offering advantages for gel tunability. Another benefit of dynamic linking strategies is the possibility of forming strong, but (thermally or chemically) reversible networks.

To probe gel dynamics across scales, it is crucial to select NP building blocks whose sizes, shapes, and interactions can be sufficiently well characterized for conditions relevant to network formation.

\section{Quantifying NP gel dynamics with photon correlation}
One of the most basic measures of single-NP dynamics is obtained by relating the NP mean-square displacement (MSD) $\left<r^2\right>$ to an observation time scale $\tau$. In a three-dimensional system exhibiting only Brownian motion, this relationship is~\cite{einstein1905, Sutherland1905}
\begin{equation}
    \langle r^2\rangle = 6D\tau
    \label{tau_scaling}
\end{equation}
where $D$ is the self-diffusion coefficient.

Direct multiple-particle-tracking techniques can assess dynamic behavior from microscopy by observing changes in particle positions across time intervals, $\tau$, similar to the procedure used in Brownian dynamics computer simulations. However, the sub-diffractive-limit size and high optical opacity of inorganic NPs are significant barriers to directly extracting dynamic behavior. Indirect characterization using scattering approaches provides an alternative for probing dynamics at sub-diffractive length scales.

\subsection{Measuring nanoparticle positions with scattering intensity}
Quasi-elastic light scattering techniques probe the intensity of scattered coherent light after interaction with a collection of particles. Scattered light creates a speckle pattern that is collected in a pixel array at the detector, with each pixel corresponding to a wavevector ${\mathbf{q}}$ with amplitude (wavenumber) $|{\mathbf{q}}|=q$. The value of $q$ is determined by the experimental setup and the properties of the incident light:
\begin{equation}
    q = \frac{4 \pi n_{\text{m}}}{\lambda_0} \sin\left(\frac{\theta}{2}\right)
    \label{q}
\end{equation}
where $n_{\text{m}}$ is the refractive index of the medium (typically assumed unity in X-ray scattering), $\lambda_0$ is the vacuum wavelength of incident light, and $\theta$ is the angle between incident beam and scattered light.

The total field of scattered coherent light $E$ at a given time $t$ is a complex number determined by the sum of the contributions of scatterers within the sample:
\begin{equation}
    E(\mathbf{q},t) \propto \sum_j a_j(q)\exp(i\mathbf{q}\cdot \mathbf{r}_j(t))
\end{equation}
where $a_j$ is the amplitude from a single scatterer, and particle position, $\mathbf{r}_j$, determines the phase of scattered light.

Importantly, the amplitude and the relationship to scattering vector depend on the ratio of the wavelength to the characteristic particle size. In visible light scattering, the wavelength exceeds characteristic NP length scales. When the wavelength is far larger than the particle diameter ($D\ll \lambda$), isotropic Rayleigh scattering from quasispherical colloids can be assumed,
\begin{equation}
    a_j \propto \left(\frac{n_j^2-n_{\text{m}}^2}{n_j^2+2n_{\text{m}}^2}\right)R^3
\end{equation}
Here, $n_j$ is the refractive index of particle $j$, $n_{\text{m}}$ is the refractive index of the medium, and $R$ is the particle radius~\cite{falke2019}. If particles have sizes similar to the wavelength, typical of microscopic colloids, the intensity exhibits a more complicated dependence on particle geometry, scattering angle, and the form of the particle's dipole relaxation described by Mie scattering theory~\cite{mie1908,fan2014}.

X-rays scatter by interacting with individual electrons rather than colloid polarizability~\cite{jeffries2021}. In X-ray scattering experiments, the wavelength is much smaller than typical NP length scales, $\lambda_0 \sim0.1$ nm. The scattering amplitude is therefore sensitive to the distribution of electrons within the colloid. The field amplitude of X-rays scattered from a randomly oriented single particle can be expressed as follows,
\begin{equation}
    a_j(q) \propto A(q)
\end{equation}
A(q) is derived from the radial electron distribution of the scatterer~\cite{li2016}: 
\begin{equation}
    A(\mathbf{q}) = \int_V \Delta \rho_e(r) e^{i\mathbf{q \cdot r}}\text{d}\mathbf{r} 
\end{equation}
where $\Delta \rho_e(r)$ is the radial electron density difference between the scatterer and the medium.

For homogeneous spherical particles with scattering radius $R$, A(q) has an analytical form, given by
\begin{equation}
    A(q) = \Delta \rho_e V_p \frac{3(\sin(qR)-qR\cos(qR))}{(qR)^3}
\end{equation}
Analytical $F(q)$ for various simple geometries are available, though its determination can be challenging for particles that are geometrically complex.

It is helpful to relate the electric field of scattered light to the intensity $I(q,t)$, which can be measured,
\begin{equation}
    I(q,t) \propto |E(q,t)|^2 = E^*(q,t)E(q,t)
\end{equation}
or equivalently
\begin{equation}
    I(q,t) \propto \sum_j\sum_ka_j(q)a_k(q) e^{i\mathbf{q}\cdot ({\mathbf{r}}_j(t)-{\mathbf{r}}_k(t))}
    \label{I_from_E}
\end{equation}

The measured intensity is dependent on the phase interference of waves scattered by all particles, and its $q$-dependent features reflect correlated density fluctuations for real-space (Bragg) length scales $d = 2\pi/q$.

Differing contrast mechanisms determine the application of visible and X-ray scattering.
Rayleigh intensity scaling of $R^6$ and high refractive index contrasts can lead to multiple scattering of the same wave. DLS is highly susceptible to multiple scattering and can distort the interpretation of measured intensity values in optically dense dispersions~\cite{ragheb2020}. X-rays scatter relatively weakly by comparison and with a forward bias, reducing the likelihood of multiply scattered light, and enabling resolution of denser colloidal NP suspensions.

In practice, the intensity is averaged to improve scattering statistics. When collecting photons at a single pixel corresponding to ${\mathbf{q}}^{\prime}$, like with standard DLS, this can only be accomplished via temporal averaging, 
\begin{equation}
    \langle I(q^{\prime},t)\rangle_T \propto \langle\sum_j\sum_ka_ja_k e^{i\mathbf{q^{\prime}}\cdot ({\mathbf{r}}_j(t)-{\mathbf{r}}_k(t))}\rangle_T
    \label{temporal_average}
\end{equation}
where $q^{\prime}=|{\mathbf{q}^{\prime}}|$, and the average is taken over a duration $T$, from $t-T/2$ to $t+T/2$.

Simultaneously measuring intensity at numerous pixels within an array, as usually done in XPCS and necessary in DDM, can resolve scattering from discrete length scales. In practice, for isotropic systems, intensities from pixels with the same wavenumber can also be averaged,
\begin{equation}
    \langle I(q,t)\rangle_{P(q)} \propto \langle \sum_j\sum_ka_ja_k e^{i\mathbf{q}\cdot ({\mathbf{r}}_j(t)-{\mathbf{r}}_k(t))}\rangle_{P(q)}
    \label{spatial_average}
\end{equation}
Here, the subscript $P(q)$ indicates an instantaneous azimuthal average over all pixels with equivalent $q=q^{\prime}=|{\mathbf{q}^{\prime}}|$ at time $t$.

Photon correlation spectroscopy leverages phase dependence of measured intensity to probe NP dynamics through analysis of intensity fluctuations over time. The precise relationship of intensity fluctuations to nanoscopic density fluctuations varies with the experimental setup, the type of dynamics, and emergent structure of scatterers.

\subsection{Simple Brownian dynamics and the one-time correlation function}

Change in the relative positions of independent scatterers causes the phase interference of the scattered waves to fluctuate~\cite{pecora1964,berne2000}, as reflected by Eq~\ref{I_from_E}. The decorrelation of the field amplitude over a lag time $\Delta t$ is captured in the normalized first-order field correlation function $g_1$,
\begin{equation}
    g_1(q,\Delta t) = \frac{\langle E^*(q,{t}) E(q,t+\Delta t)\rangle_{t}}{\langle| E(q,{t})|^2\rangle_{t}}
    \label{field_corr}
\end{equation}
with limits of $g_1(q,0) = 1$, where particle positions have not changed, and $g_1(q,\infty) =0$, where particle positions no longer resemble those at time ${t}$. Here, the average is taken over time $t$ within a period for which the dynamics are stationary. The field correlation function can alternatively be expressed as a normalized intermediate scattering function~\cite{leheny2012}
\begin{equation}
    g_1(q,\Delta t) = \frac{F(q,\Delta t)}{F(q,0)}
\end{equation}
where $F(q,\Delta t)$ is the intermediate scattering function, and $F(q,0)=S(q)$ is the static structure factor.

When scatterers diffuse independently, such as during Brownian motion, the distribution of particle displacements is Gaussian, and $g_1(q,\Delta t)$ decays exponentially,~\cite{ferreira2020}
\begin{equation}
    g_1(q,\Delta t) = \exp\left(- \frac{\Delta t}{\tau(q)}\right)
    \label{g1_gauss}
\end{equation}
The wavenumber dependence of the diffusion rate, $\Gamma(q) = 1/\tau(q)$ of non-interacting Brownian diffusers can then be directly written as the reciprocal-space analog of Eq~\ref{tau_scaling}~\cite{stetefeld2016},
\begin{equation}
    \Gamma(q) = Dq^{2}
    \label{reciprocal_tau}
\end{equation}
enabling the following convenient analytical representation of Eq~\ref{g1_gauss},
\begin{equation}
    g_1(q,\Delta t) = \exp(-Dq^2\Delta t)
\end{equation}
The spatiotemporal representation of the field decorrelation for scatterers undergoing Brownian motion is shown in Figure~\ref{g1_char}b. If interparticle interactions are present, the self-diffusion coefficient is replaced with a time- and wavenumber-dependent diffusivity, $D(q,t)$~\cite{segre1997dynamics}.

Relating observable intensity fluctuations to the underlying decorrelation of scatterer positions indirectly quantifies NP motion. The one-time normalized intensity correlation function $g_2$ evaluates similarity in average scattering intensity at a reference time $t$ and after a lag time $\Delta t$:
\begin{equation}
    g_2(q,\Delta t) = \frac{\left\langle I(q,t)I(q,t+\Delta t )\right\rangle_{P(q),t}}{\left\langle I(q,t)\right\rangle^2_{P(q),t}}
        \label{ICF}
\end{equation}
with perfectly coherent light yielding limits of $g_2(q,0) = 2$, arising from Gaussian statistics of a fully developed intensity field, $\langle I^2 \rangle = 2 \langle I\rangle^2$, and $g_2(q,\infty)=1$, where the intensity observations are independent~\cite{siegert1943}.

The intensity correlation is calculated using the temporal ensemble average of intensity fluctuations considering $N_t$ total reference times and a particular lag time $\Delta t$:
\begin{equation}
    \langle I(q,t) I(q,t+\Delta t)\rangle_{P(q),t} = \frac{1}{N_t}\sum_{j=1}^{N_t}\langle I(q,t_j)I(q,t_j+\Delta t)\rangle_{P(q),T}
\end{equation}
where intensities at times $t$ are averaged over both pixels and finite temporal windows, $T$, around $t$.

While the temporal binning at reference times can be adjusted to improve the scattering statistics, only the spatial averaging is strictly necessary to obtain average intensity in speckle array measurements. Later in this section, we discuss the numerous advantages of wavenumber-resolved dynamics over single pixel techniques, which assume equivalence of temporal and spatial averaging.

Assuming Gaussian electric field distributions and ergodicity, the intensity correlation function can be related to the field correlation, $g_1$, through the Siegert relationship~\cite{siegert1943}:
\begin{equation}
        g_2(q,\Delta t) = 1+\beta |g_1(q,\Delta t)|^2
        \label{field_correlation}
\end{equation}
where $\beta$ describes the speckle contrast that mostly depends on the coherence of the incident light.

\subsection{Stretched and compressed dynamics}
As interparticle interactions begin to influence NP diffusion, it becomes necessary to account for anomalous motion deviating from assumed Gaussian displacements. The Kohlrausch-Williams-Watts (KWW)~\cite{kohlrausch1854,williams1970} relation captures the breadth of relaxation processes by stretching the relaxation distribution with a single exponent, $\gamma$:
\begin{equation}
    g_1(q,\Delta t) = \exp\left[-\left(\frac{\Delta t}{\tau (q)}\right)^{\gamma(q)}\right]
    \label{kww}
\end{equation}
If dynamic relaxations are heterogeneous, then $\gamma<1$, and the intermediate scattering function is stretched across the lag time. Stress and strain propagation, structural collapse, or cooperative dynamics exhibit a characteristic $\gamma>1$, reflecting sharper homogeneous decays~\cite{lehmkuhler2020}. The different resulting curvatures of the normalized field correlation function are shown in Figure~\ref{g1_char}A.

Anomalous dynamics alter scaling of relaxation time with wavenumber to reflect non-Gaussian particle displacements:
\begin{equation}
    \tau(q) \propto q^{-p}
    \label{tau_q}
\end{equation}
Values of $p>2$ indicate relaxation times consistent with \textit{subdiffusive} motion. Superdiffusive motion produces $p<2$, with the limit of ballistic motion corresponding to $p=1$. This type of analysis can be extended to describe other models for stochastic processes which may apply for specific systems depending on the nature of the NPs, their interactions, and the relevant dynamics~\cite{metzler2014}.

In a system with a single, scale-independent distribution of dynamic processes, $\gamma$ is constant across $q$. Gels with length-scale-heterogeneous dynamics may necessitate $q$-dependent assessment of the scaling exponent. Comparing $\gamma (q)$ to the spatial scaling of dynamics in Eq~\ref{tau_q} allows characterization of dynamic processes through both the temporal distribution of relaxation times and their dependence on observed length scale~\cite{jain2022,ishii2021}. This is useful for assessing the emergence of scale-dependent dynamic processes as NP gels undergo initial transitions from fluid dispersion to gel.

\subsection{Nonergodicity and structural influence on measured dynamics}
\begin{figure}[hbt]
    \centering
    \includegraphics[width = 0.5 \textwidth]{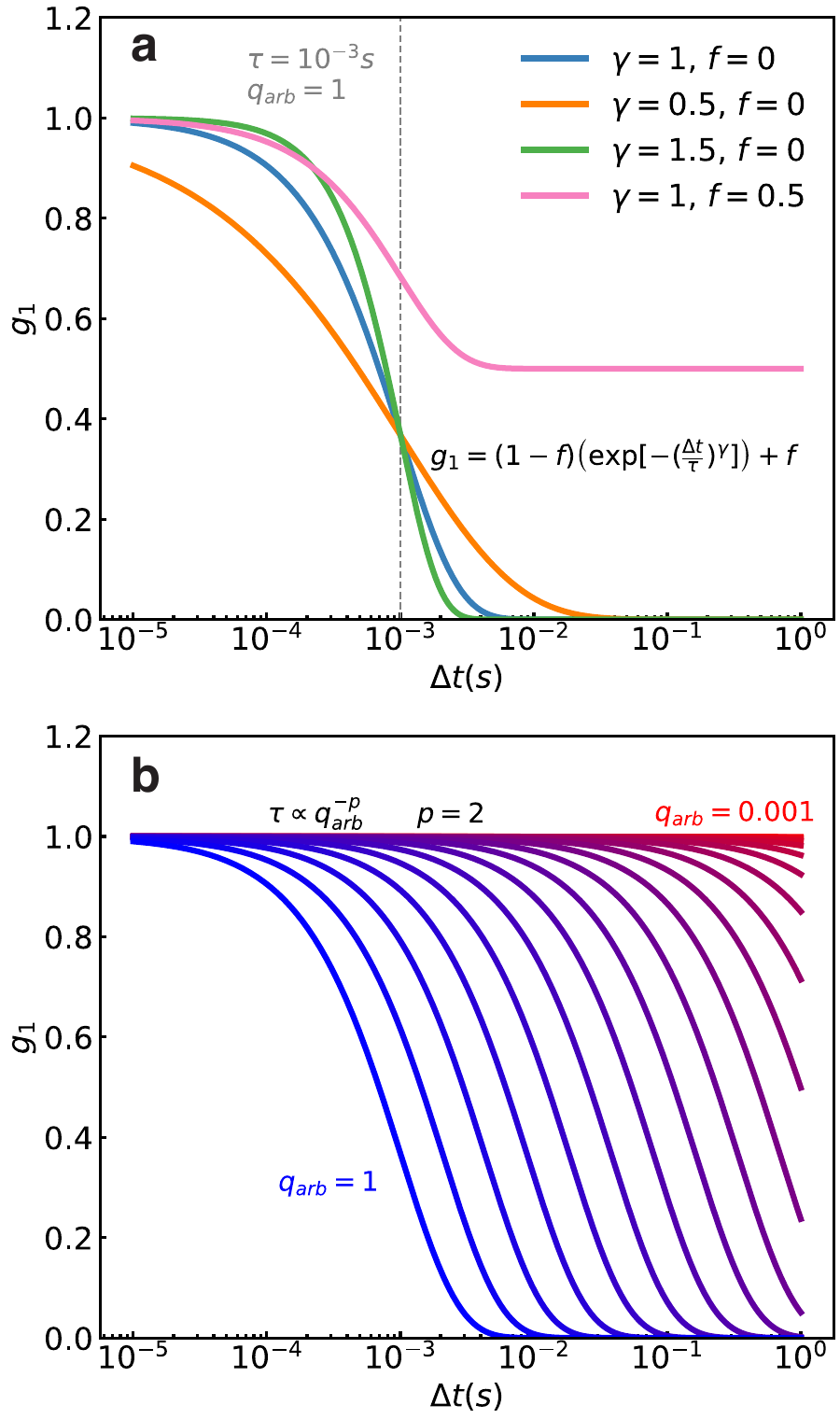}
    \caption{Calculated intermediate scattering functions for (a) relaxation time, $\tau = 10^{-3}s$ at arbitrary scattering vector, $q_{arb} =1$, varied KWW exponent, $\gamma$, and nonergodicity factor, $f$ and (b) Gaussian diffusion with relaxation time, $\tau = 10^{-3}s$ scaled with scattering vectors ranging from $q_{arb} = 0.001$ to $q_{arb} = 1$.}
    \label{g1_char}
\end{figure}

Averaging over many speckle patterns of equal $q$ can be advantageous when evaluating NP gel dynamics. First, when the scattering volume is large enough, pixel averaging effectively integrates the correlations of many independent sub-ensembles, allowing one to decouple spatial and temporal sampling. This is particularly relevant for gels, where a fraction of the NPs become spatially localized, meaning that the time-averaged intensity of a single pixel of a given $q$ may no longer be equivalent to the ensemble average at that wavenumber, i.e.,
\begin{equation}
    \langle I(q,t) \rangle_{P(q),t} \neq \langle I(q,t) \rangle_{t}
\end{equation}
This phenomenon, known as \textit{nonergodicity}, is a characteristic of NP gels and glasses, where multi-speckle correlation for tracking emergent localized dynamics is commonly applied~\cite{bohidar2000}. At length scales beyond those fully explored by all NPs (i.e., at sufficiently low q), nonergodic systems exhibit a characteristic plateau in the time-averaged field correlation, reflecting persistent structural correlations~\cite{pusey1989}.  To account for this plateau behavior, the first-order correlation is modified to include a nonergodicity factor, $f(q)$, (see pink curve in Figure~\ref{g1_char}a)~~\cite{cho2020},
\begin{equation}
    g_1(q,\Delta t) = (1-f(q))\exp\left[-\left(\frac{\Delta t}{\tau (q)}\right)^{\gamma (q)}\right] +f(q)
    \label{nonergodic_field}
\end{equation}
This expression is then used together with Eq~\ref{field_correlation} to describe the spatiotemporal evolution, taking into account the appropriate limits
of the normalized field and intensity correlations:  $g_1(q,\infty) = f(q)$ and $g_2(q,\infty) = 1+\beta f(q)^2$, respectively.
Tracking the evolution of $f(q)$ provides a quantitative measure of network formation in aging and nonergodic relaxation~\cite{bohidar2000,cho2020, jain2022, ofosu2022, chen2023}.
Figure~\ref{g1_char}a displays the calculated normalized field correlation for anomalous and localized dynamics at an arbitrary scattering vector and characteristic relaxation time, $\tau = 10^{-3}s$.

Fits of data acquired from XPCS to a model for the intermediate scattering function can be used to help identify the fraction of localized particles within the scattering volume and their vibrational characteristics~\cite{guo2011}. A Debye-Waller type relationship is typically used to characterize localized modes,
\begin{equation}
    f(q) = f_0 \exp\left(-\frac{q^2 r_{loc}^2}{6}\right)
    \label{debye_waller}
\end{equation}
where $f(0)$ is the fraction of particles localized to a characteristic length scale, $r_{loc}$. 

Analysis of experimental data of NP gels can also account for dynamics \textit{faster} than the minimum accessible lag time. The decorrelation that occurs before the observation window features as an apparent decrease in the contrast~\cite{jain2020}:
\begin{equation}
    g_2(q,\Delta t) = 1+\frac{\beta}{b_0}|g_1|^2
\end{equation}
where the short-lag-time plateau at $g_2(q,0) - 1$ is lower than the conventional value by a factor of $b_0$.
In $g_2$ analysis similar to Eq~\ref{debye_waller}, the contrast ratio can be used to determine the dynamic heterogeneity at shorter time scales. If the faster process is diffusive, this can be accomplished using the following expression~\cite{jain2020},
\begin{equation}
    \beta/b_0 = n_0^2 \exp\left(-\frac{q^2 r_{\text{loc}}^2}{3} \right)
\end{equation}
where $n_0$ is the fraction of particles with unresolved short-time relaxations. 

XPCS can resolve nanoscopic interparticle structural correlations as clusters and gel networks develop. In dense fluids, particle positions at structurally correlated length scales decorrelate more slowly than at unstructured length scales. This phenomenon is known as \textit{de Gennes narrowing}, and results in a modified spatial scaling of relaxation time scales~\cite{kellouai_gennes_2024}:
\begin{equation}
    \tau(q) = \frac{S(q)}{D_0q^2}
\end{equation}
where the relaxation time scales with the ratio of interparticle structure factor, $S(q)$, to the particle diffusivity, $D_0$. 

If NP gels have coarsened and exhibit significant interparticle structure, the localization length scales of dynamic modes must also account for de Gennes narrowing, modifying Eq~\ref{debye_waller}~\cite{leheny2012}:
\begin{equation}
    f(q) = f_0 \exp\left(-\frac{q^2 r_{\text{loc}}^2}{6S(q)}\right)
\end{equation}
with the nonergodicity now reflecting the localization of dynamics relative to the static structure factor, $S(q)$.

In some cases, localized dynamic modes observed within fractal gel networks can be attributed to elastic vibrations. The resulting $q$-independent relaxation times have been described with~\cite{cho2020}:
\begin{equation}
    \tau \sim \phi^{(\alpha-1)/(3-d_f)}
\end{equation}
where measured relaxation time depends on the scatterer volume fraction, $\phi$, an elasticity exponent derived from the spring constant between bonded particles, $\alpha$, and fractal dimension, $d_f$.

These analyses provide a framework for quantifying how structural arrest and dynamical heterogeneity propagate across length scales in NP gels. Existing analyses largely treat these phenomena independently and are often limited to pre-formed gel networks. Critical questions remain about how anomalous diffusion, nonergodicity, and dynamics of NP gel structures actively emerge and interact during gelation. Resolving how nanoscopic arrest and network dynamics manifest across gel length scales can elucidate the nanoscopic origins of bulk properties such as network rigidity. The full spectrum of spatiotemporal dynamics accessible via speckle pattern correlation is also necessary to determine how gel design principles can be used to tune gelation kinetics and network mechanics, and to develop models to describe the codependency of emergent dynamical features.

\subsection{Multiple dynamic processes}
Intensity fluctuations can also decorrelate in multiple steps when dynamic processes with sufficiently different mean characteristic relaxation times occur within the same lag time window~\cite{lienard2022}. Multi-step relaxations produce composite intermediate scattering functions of individual relaxations weighted by the fraction of scatterers participating in each process. For a system of $N$ diffusive processes, $g_1$ can be expressed as~\cite{shukla2004, jain2022}:
\begin{equation}
    g_1 = \sum_i^N A_i \exp\left(\frac{-\Delta t}{\tau_i}\right)
\end{equation}
where $A_i$ is the fractional amplitude of scatterers with mean relaxation time, $\tau_i$. If the processes are anomalous or nonergodic, each process may alternatively resemble Eq~\ref{kww} or Eq~\ref{nonergodic_field}, respectively. 

\subsection{Nonequilibrium dynamics and the two-time correlation function}
Another advantage of speckle averaging is the ability to generalize correlations using instantaneous intensity averages. When dynamics change during scattering measurements, the one-time correlation function can no longer describe decorrelation with ensembles obtained by averaging over all measurement times. Evolving dynamic behavior can instead be characterized with the more general two-time correlation (TTC) ~\cite{madsen2010,bikondoa2017,ragulskaya2024analysis}:
\begin{equation}
    {\text{Corr}}(q,t_1,t_2) = \frac{\langle I(q,t_1)I(q,t_2)\rangle_{P(q)}}{\langle I(q,t_1)\rangle_{P(q)}\langle I(q,t_2) \rangle _{P(q)}}
    \label{TTC}
\end{equation}
where $t_1$ and $t_2$ are speckle-averaged intensity values at different measurement times.
The TTC consequently forms a two-dimensional array of intensity values symmetric along the diagonal $t_1 = t_2$. 

It is common to relate the TTC and Siegert relation by temporally averaging intensity fluctuations around constant sample age, $t_{\text{age}} = (t_1+t_2)/2$. Averaged sections of the TTC perpendicular to the diagonal at $t_{\text{age}}$ can then be treated as quasiequilibrium one-time intensity correlation functions: 
\begin{equation}
    g_{2,t_{\text{age}}}(q,\Delta t) = 1 + \beta|g_{1,t_{\text{age}}}(q, \Delta t)|^2
\end{equation}
where $\Delta t = |t_1-t_2|$ along the perpendicular slice. 

When the dynamics do not change during measurement, averaging across all values of $t_{\text{age}}$ returns the same $g_2$ as a standard one-time intensity correlation function. TTC of speckle patterns can also be used for evaluate higher order correlations describing heterogeneous or nonequilibrium dynamics common in aging NP gels~\cite{bikondoa2017} or externally driven systems~\cite{he2024,horwath2024}. 

As we outline in subsequent sections, TTC analysis of measurements at upgraded XPCS beam lines and in novel in situ experimental setups is becoming a popular approach for understanding the complex multiscale dynamics characteristic of colloidal NP gels.

\section{The expanding nanoparticle gel dynamics toolbox}
Structural length scales and relaxation time scales ranging several orders of magnitude during NP gelation makes accessing the full spatiotemporal dynamic spectrum in these systems especially challenging. Significant improvements in XPCS technology have expanded typical measurements to relaxations ranging from $10^{-6}$ to $10^{3}$~s time scales and length scales from NP dimensions to microns.

\begin{figure*}[ht!]
    \centering
    \includegraphics[width = \textwidth]{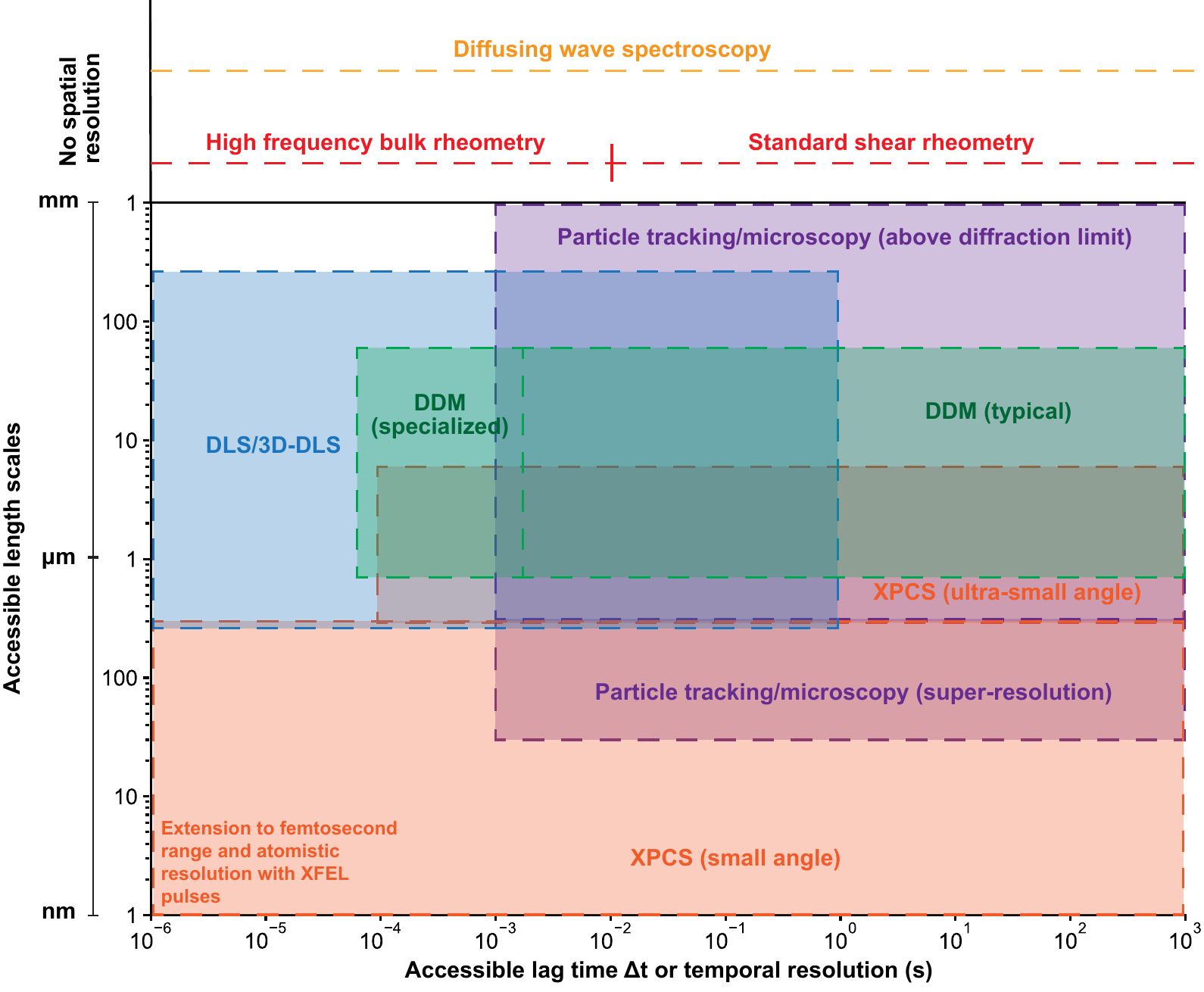}
    \caption{Approximate spatiotemporal limits for individual experimental dynamic characterization techniques. Regions for respective techniques may have reduced practical range depending on sample properties and experimental configuration.}
    \label{technique_spatiotemporal}
\end{figure*}

\begin{figure*}[ht!]
    \centering
    \includegraphics[width = \textwidth]{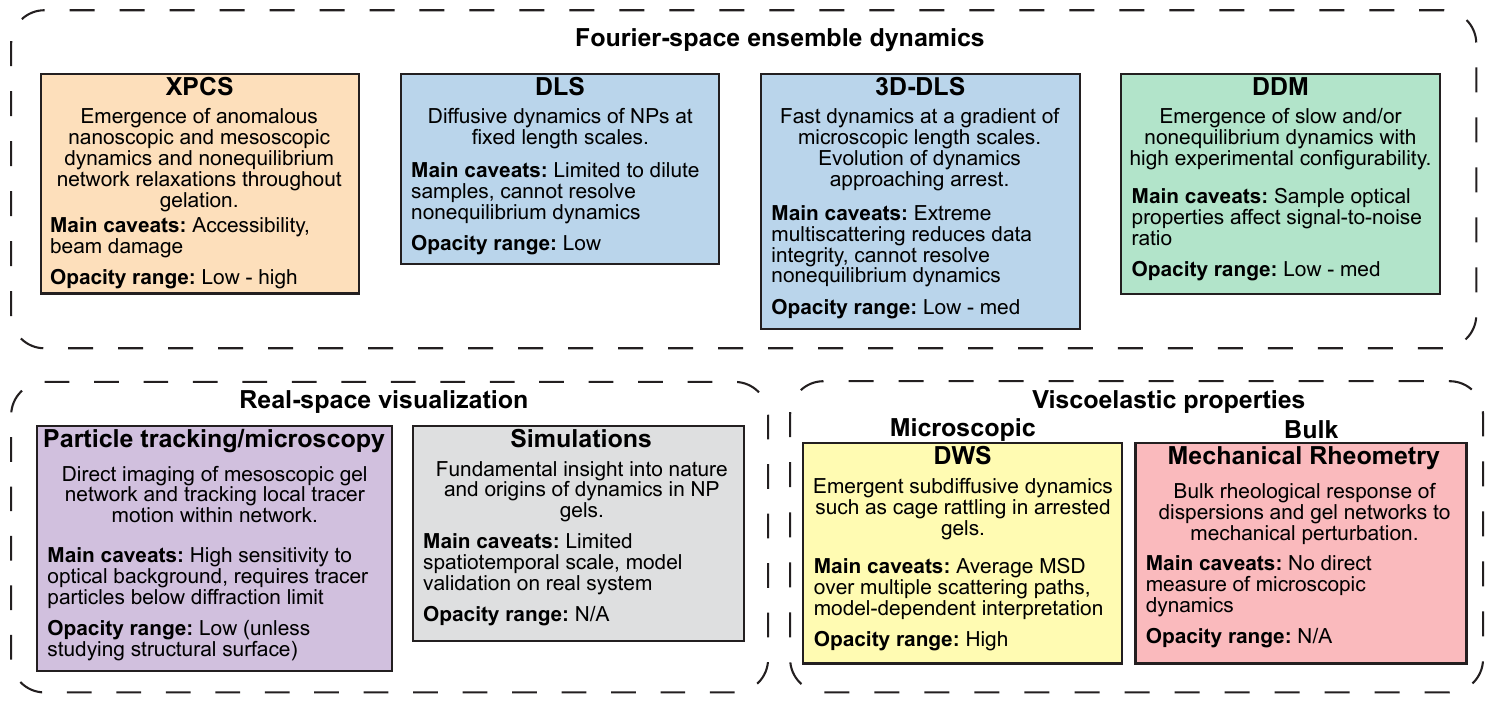}
    \caption{Comparison of major techniques used to characterize soft matter dynamics with respective potential use to characterize NP gels, limitations, and applicable optical density.}

    \label{technique_overview}

\end{figure*}

While XPCS is particularly well-suited for characterizing dynamics in NP gels, thorough characterization of emergent NP network evolution requires insights from dynamics at large length scales, real-space structural properties, and network mechanics. Here we discuss how state-of-the-art light scattering, microscopy, particle tracking, rheology, and simulations can supplement XPCS by extending accessible length scales (Figure~\ref{technique_spatiotemporal}) or providing accessible routes for lab-scale experiments to resolve relevant NP dynamics (Figure~\ref{technique_overview}).

\subsection{X-ray photon correlation spectroscopy}
\begin{figure*}[ht]
    \centering
    \includegraphics[width = \textwidth]{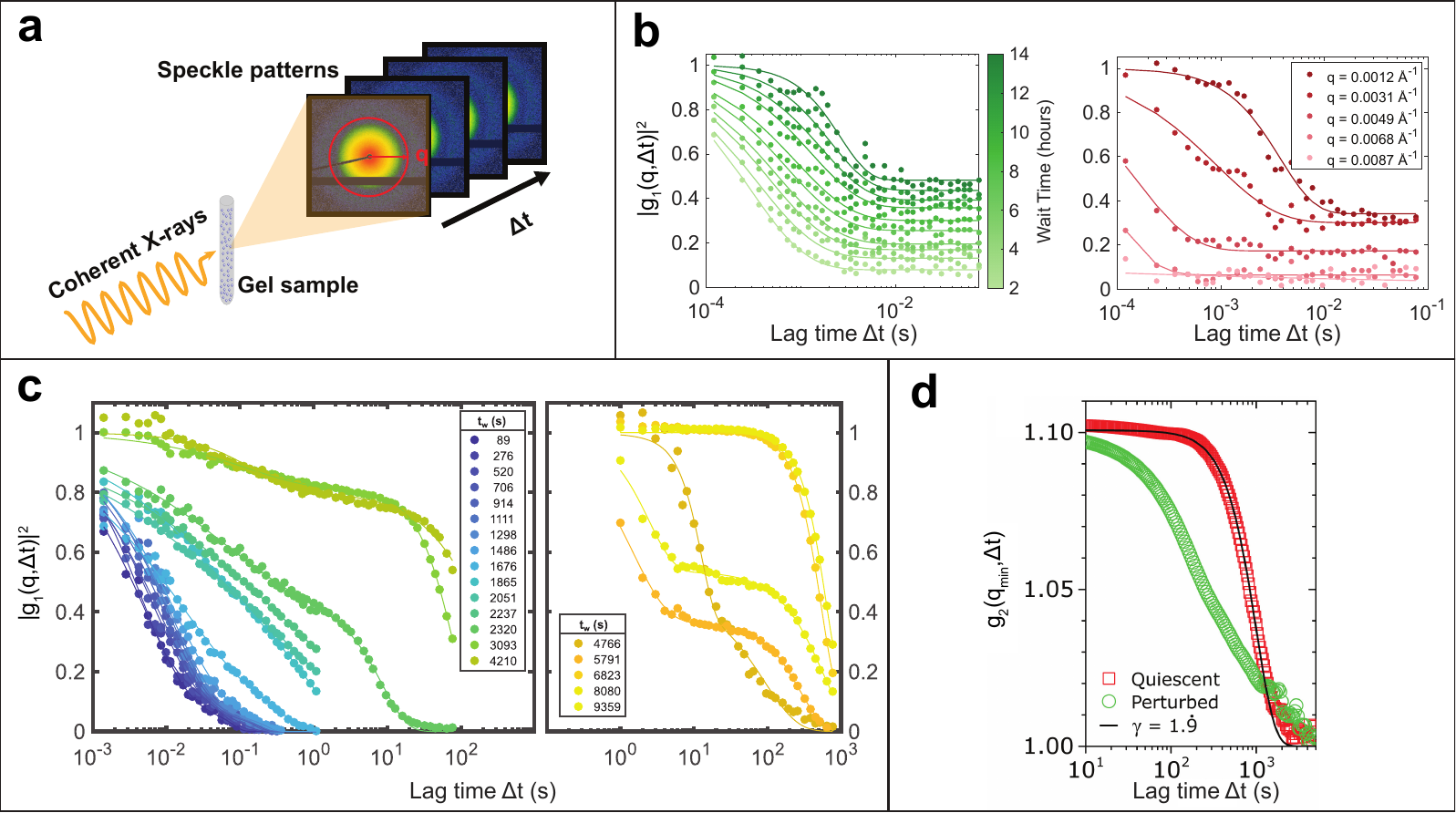}
    \caption{(a) Schematic of XPCS. Adapted from~\cite{schroer2019}. (b) Evolution of single stretched correlation functions over time at fixed $q = 0.0031$~\AA$^{-1}$ (left) and over scattering vector at a fixed wait time of 8.3 hours (right) for molecularly linked tin-doped indium oxide nanocrystals. Adapted from~\cite{ofosu2025}, licensed under CC 4.0. (c) Emergence of multi-step correlation function at fixed sample spot (left) and varied sample spots (right) for PEGylated gold NPs. Adapted from~\cite{jain2022}, licensed under CC 4.0. (d) Ballistic correlation function (red) and correlation function for mechanically perturbed gel network (green) for iron oxide NPs gelled by linking with 4-arm PEG. Reproduced from~\cite{song2022},~licensed under CC 4.0.}
    \label{XPCS_figure}

\end{figure*}

Inorganic NPs have high electron density owing to heavy element compositions and are well-suited for X-ray scattering~\cite{ahmad2023}. Despite being good scatterers, low X-ray scattering cross sections and short sample path lengths in capillaries or sandwich cells minimize multiple scattering effects in dense NP samples, enabling characterization of all NP gelation stages. Small angle X-ray scattering is sensitive to nanoscopic structural correlations, making XPCS at small angles well suited for characterizing dynamics from single NP motion to dynamics coupled across different structural scales in NP gels. As described in Section III, dynamics at different nanoscopic structural length scales are extracted by assessing fluctuations at speckle pattern radii corresponding to scattering vector, $q$, as demonstrated in Figure~\ref{XPCS_figure}A.

Analysis of resulting correlation functions has been used to characterize emergent anomalous dynamics and $q$-scaling (Figure~\ref{XPCS_figure}b), including multiple dynamic processes (Figure~\ref{XPCS_figure}c) and how collective relaxations are heterogeneously stretched during mechanical perturbation (Figure~\ref{XPCS_figure}d).
Azimuthal averaging of speckle patterns has been used to quantify dynamic anisotropy in aging NP gels~\cite{jain2020}, and metallic glasses~\cite{wang2024}.

XPCS intensity directly affects the signal-to-noise ratio in extracted $g_2$ functions. In low contrast NP gels (e.g. with smaller particle size or lighter elements in the NP core), limited scattering statistics may require longer averaging times for clean speckle patterns. Conversely, high intensities can also damage X-ray absorbing NP gels. High beam coherence can mitigate this sensitivity by generating better speckle statistics at lower intensities. 

X-rays must be (at least partially) spatially coherent to resolve speckle patterns. Achieving highly coherent X-rays requires extensive spatiotemporal filtration, which significantly attenuates the incident intensity~\cite{nogales2016}. High intensity coherent X-rays are only currently available at synchrotron radiation sources and X-ray free electron lasers. Until recently, XPCS was only used at a select few facilities, including APS, ESRF, PETRA III, and NSLS-II, but recent advances in light sources at synchrotron radiation and X-ray free electron laser sources have significantly expanded XPCS capabilities. The main limitation of XPCS is ultimately access to high-intensity, coherent light sources. 

Fourth-generation light sources and better detectors offer high spatiotemporal resolution with high X-ray flux and more speckle resolution. While XFEL-based XPCS dramatically extends spatiotemporal limits, synchrotron sources continue to advance rapidly. Thanks to its particular signal-to-noise ratio~\cite{falus2006, lehmkuhler2021}, XPCS experiments can access up to $n^2$ shorter correlation times upon an increase of a factor of $n$ of brilliance. The upgrade to a fourth generation light source results, for almost all facilities, in a brilliance gain in the range of 100~\cite{perakis2020, lehmkuhler2021}, resulting in access to $10^4$ times shorter relaxation times with the same signal-to-noise ratio. As a consequence, XPCS will routinely enable access to microsecond time scales for soft matter samples. These advances in the X-ray sources further demand development of fast detectors. Recently, several studies have demonstrated XPCS at this time scale at APS, PETRA III and ESRF-EBS following different detector developments~\cite{zhang2018, jo2021, chushkin2025}. 

Ultra-small-angle configurations at APS and ESRF reach scattering vectors $q=10^{-4}$ \AA$^{-1}$ to probe micron-scale structural fluctuations~\cite{ilavsky2018,chevremont2024}. Wide-angle configurations extend up to $q=6.5$ \AA$^{-1}$~\cite{allen2023}, resolving scales well below typical NP dimensions. Recent work at the European XFEL has demonstrated MHz XPCS on radiation-sensitive soft matter systems \cite{lehmkuhler2020,reiser2022,dallari2024,girelli2025}. In addition, FEL sources demonstrated femtosecond relaxation dynamics with atomic resolution, such as in XPCS measurements of metallic glasses~\cite{fujita2025}. These capabilities collectively span length scales from the nanometer scale to scales $>600$ nm and time scales from femtoseconds to hours, positioning XPCS as a powerful technique for probing multi-scale dynamics in NP gels.

In addition to dynamics, XPCS enables measurement of static nanoscopic structure by appropriately averaging speckle patterns obtained during dynamic measurements. Ensemble-averaged XPCS speckle patterns can be compared to conventional small-angle X-ray scattering (SAXS), widely used to quantify the equilibrium structure of nanoscale materials~\cite{li2016}. In ergodic systems, this is accomplished by averaging intensity contributions from many decorrelated subensembles. For a dispersion of $N$ identical particles, the relevant relation is given by
\begin{equation}
    \langle I(q)\rangle_{P(q),t\gg t'} \propto A(q)^2S(q)
\end{equation}
where the intensity is averaged over times far greater than the persistence time of correlated particle positions $t'$, allowing for the calculation of the ensemble-average of interparticle correlations per Eq~\ref{I_from_E}.
For nonergodic samples, sampling of many spatial subensembles can be assumed to incoherently add to the average speckle pattern. This requires rotating the sample, moving the incident beam, or expanding the scattering volume with long path lengths or large beam spot sizes~\cite{shah_scattering_2003}.

The ability to probe nonequilibrium structural relaxation from the nanoscale up to the microscale has made XPCS an indispensable technique for characterizing colloidal NP dynamics. However, the necessity of X-ray facilities such as synchrotron radiation sources or XFELs to generate coherent X-rays limits widespread adoption. The indirect nature of light scattering can also make it difficult to link observed dynamic phenomena to real-space gel structure. Complementary visible light scattering techniques can access additional spatiotemporal ranges or enable rapid measurements, while direct visualization of gel networks would structurally contextualize dynamics of gels in real-space.  

\subsection{Visible light scattering}
Visible and near infrared (NIR) dynamic light scattering (DLS) provides an accessible route for probing NP dynamics with common coherent laser sources. Generally, DLS excels at probing relative scatterer motion at length scales beyond the local nanoscopic structural fluctuations assessed with XPCS. CCD cameras can be used to provide multi-speckle averaged intensity correlation function~\cite{cipelletti1999, golde2013, franke2014}, but multiple scattering practically limits its application in inorganic systems dense enough to exhibit nonergodic behavior. DLS is realistically best-suited to characterize anomalous dynamics resulting from long-range interactions between NPs or small clusters unless measures are taken to minimize scattering path lengths.

\subsubsection{Standard single-speckle dynamic light scattering}
Standard DLS employs a single detector fixed at a specific angle to measure correlations related to the diffusive motion of NPs and microscopic aggregates at a single $q$ value.

Accessible lag times can vary by instrument, but DLS can reach correlation lag time ranges from approximately $10^{-6}$ to 1~s and are optimized for measuring fast diffusion of small NPs up to slower diffusion of disperse NP clusters. 

The main advantage of DLS is \textit{accessibility}. Commercial instruments are common and have been extensively used to quantify the diffusive dynamics of inorganic NPs across a range of sizes, surface chemistry, and solvent environments~\cite{leong2018,mozaffari2019,zheng2016,rodriguez-loya2023}. DLS provides a reliable baseline for evaluating the fundamental dynamic properties of primary NPs before incorporation into gels. Its simplicity and rapid acquisition time make it useful for high throughput assessment of how primary particle modifications or solvent environments influence NP mobility in dispersion.

\subsubsection{3D dynamic light scattering}
Three-dimensional dynamic light scattering (3D-DLS) employs multiple detectors to cross-correlate scattered intensity and correct for multiple scattering effects~\cite{schatzel1991}. Using static or mobile detectors at varied angles probes multiple $q$-ranges, while cross-correlation enables the measurement of higher NP volume fractions~\cite{pusey1999,block2010}.

Modulated 3D-DLS has accurately resolved dynamics of dispersions with transmission as low as 1\%~\cite{pal2022}. To further extend the turbidity range for 3D-measurements, the sample path length can be changed using methods such as measuring through the corner of a square sample cell~\cite{zotero-item-4363}.

State-of-the-art commercial instruments reach a wide range of scattering angles from approximately $10^\circ$ to $150^\circ$ using mobile detectors on a goniometer arm~\cite{zotero-item-4362}. For a reference, a 633 nm laser probing particles dispersed in water corresponds to scattering vectors spanning $q\approx2.6\times10^{-2}$ nm$^{-1}$ to $2.3 \times10^{-3}$ nm$^{-1}$ or length scales of 250 nm to 2700 nm. 3D-DLS can access dynamic time scales similar to standard DLS, ranging from microsecond to single-second regimes.

3D-DLS is typically used to measure stationary systems due to its use of single speckle detectors. While uncommon, nonergodic turbid media has been investigated using methods proposed to extract ensemble averaging from 3D-DLS using the dynamic structure factor~\cite{haro-perez2011,pusey1989} or physically moving the sample to measure multiple sub-ensembles~\cite{medebach2008}. However, dynamics that evolve at time scales faster than manipulation of sample or detector position or excessive multiple scattering may prevent accurate ensemble averaging.

3D-DLS offers a powerful route to probe primary particle dynamics at volume fractions beyond standard DLS. Access to a broad $q$-range further allows the analysis of both short-range steric or ligand-mediated interactions and long-range electrostatic dynamic effects. 3D-DLS serves as a natural complement to XPCS by extending dynamic characterization into micron length scales and may be useful to investigate length-scale-dependent NP or cluster dynamics across a wider range of volume fractions than standard DLS.

\subsubsection{Diffusing wave spectroscopy}
Diffusing wave spectroscopy (DWS) is a correlation spectroscopy that can be viewed as an extension of DLS designed specifically for strongly scattering attractive glassy colloidal materials~\cite{pine1988}. Unlike DLS, which characterizes dynamic fluctuations based on single scattering events, DWS treats light as a wave diffusing through multiple scattering paths in optically dense or opaque materials. DWS can also be used with multispeckle detectors to evaluate heterogeneous dynamics.~\cite{ju2022}.

DWS measures non-diffusive relaxations of probe particles arising from local hydrodynamic interactions in dense suspensions~\cite{weitz1989}. Through modeling described elsewhere~\cite{orsi2025,mancebo2024}, the MSD of scattering particles is obtained at length scales determined by depth of light penetration. The MSD can then be related to the local viscoelastic properties of the network around scatterers. 

Multispeckle DWS (MSDWS) has been used to study soft systems with nonequilibrium dynamics. Unlike XPCS, MSDWS does not average over discrete $q$-range speckles. It instead captures speckle-averaged, multiply scattered light near a single scattering vector to track heterogeneous dynamics at a single penetration depth~\cite{zakharov2006,xu2021}. By averaging intensity fluctuations across many speckles, MS-DWS preserves autocorrelation information that would otherwise be lost~\cite{ju2022}.

Modern implementations of DWS cover a wide range of measurement time scales, extending up to days of acquisition~\cite{zakharov2006}. While two-cell DWS has been used to measure lag times down to 10 ns~\cite{viasnoff2002}, data noise makes resolving fast relaxations difficult. Recent work combining two-cell and echo DWS has proposed accurate autocorrelation function resolution across lag times ranging from $10^{-8}$ to $10^2$ s~\cite{helfer2025}.

DWS has been used extensively to characterize the microrheology of gels and composites with particle sizes $>100$ nm. While not yet commonly applied to NP gel networks, it could be envisioned to gain traction in microrheological characterization of local network properties, such as caging or fractal vibrations.

\subsection{Differential dynamic microscopy}
DDM analyzes pixel-intensity correlations in Fourier-transformed images from time-lapse optical microscopy, providing ensemble-resolved dynamic information for particles smaller than the diffraction limit. DDM owes its wide use in colloidal dynamics to its experimental versatility. Bright field DDM has been used to characterize NP Brownian dynamics~\cite{cerbino2008}, arrest of NP gel coarsening~\cite{gao2015microdynamics},
polymer-entangled NPs~\cite{shokeen2017}, and microscopically confined NPs~\cite{hitimana2022}. By using different microscope setups, DDM can be used in polarized~\cite{nixon-luke2020}, confocal, fluorescence~\cite{lu2012}, and dark field configurations~\cite{bayles2016} to characterize a wide variety of NP.

Multiple scattering effects can be mitigated with micron-scale thickness optical cells, using partially coherent light~\cite{giavazzi2014}, or measuring through a pinhole to reject out-of-plane light~\cite{lu2012}.

The $q$-range and temporal resolution of DDM strongly depend on experimental configuration. Accessible scattering vectors are generally determined by the optical magnification, aperture, illumination wavelength, and image pixel size~\cite{lattuada2025}. Typical use can access scattering vectors in the approximate magnitude ranges $0.1~\mu {\text{m}}^{-1}<q<10~\mu {\text{m}}^{-1}$~\cite{bayles2016,martinez2012,lattuada2025,schaub2025}, corresponding to length scales from sub-micron to tens-of-microns.

Accessible lag time is primarily set by the image acquisition rate and averaging method. The temporal resolution is determined by the camera frame rate, but new techniques improve temporal resolution beyond the shutter speed by alternating illumination colors (3 ms)~\cite{you2021}, cross-correlating DDM speckle patterns across multiple cameras (125 $\mu$s)~\cite{arko2019}, or pulsing LED illumination (80 $\mu$s)~\cite{schaub2025}.

DDM's recent increase in utility bodes well for its application to NP gel systems. The multi-modal possibilities make this technique robust for a wide variety of systems and a viable platform developing specified instruments to meet different experimental requirements.

\subsection{Direct particle tracking with microscopy}
Particle tracking describes a class of dynamic characterization localizing individual particles to directly extract MSD. While large colloids can be directly seen by standard wide field microscopy, the diffraction limit of visible light ($\sim250$~nm) fundamentally limits resolution at the nanoscale~\cite{macdonald2015}. 

Confocal fluorescent microscopy of fluorescent NPs has been used to overcome this limit to characterize microrheological properties of gels. The dynamics of quantum dots with varying interparticle stickiness have been used to characterize interactions between NPs and polymer melts~\cite{park2020}. The size-dependent dynamics of fluorescent tracers in supercooled matrices of $>100$ nm polystyrene NPs have also recently been investigated using fluorescence microscopy~\cite{edimeh2025}.

Some techniques use the intrinsic optical properties of inorganic NPs to localize their position. Dark field plasmon excitation microscopy has been employed to resolve plasmonic NPs and NP clusters on glass~\cite{abadie2024}, in buffer solution~\cite{liu2014}, air, water, and oil~\cite{curry2006}, and in microfluidic flow cells~\cite{sikes2022}. Dark field localization of plasmonic NPs is a quickly growing field and may be useful for characterizing plasmonic NP interaction with surrounding soft media~\cite{shi2025}.

Nanoparticle tracking analysis (NTA) is a recently commercialized technique that uses an angled camera to detect high intensity scattered light from 10-100 nm particles. While currently primarily used for diffusion sizing of biomolecules, NTA has been used to observe diffusive behavior of metal NPs~\cite{maguire2017,hole2013}. Advanced extensions of NTA use the interferometric scattering of collimated light for three-dimensional tracking of inorganic colloidal NPs in solution with as high as microsecond resolution~\cite{kasaian2024}.

At the cutting edge of microscopy research, techniques such as 3D Single-Molecule Active-feedback Real-time Tracking (3D-SMART) microscopy are paving the way for tracking NP trajectories in dense gel media~\cite{lin2025}. 3D-SMART has shown localization precision down to 30 nm with an acquisition time of a single millisecond. While promising, these techniques require complex experimental setups and are not yet widely used for studying dynamics in dense colloidal NP samples.

Microscopy techniques generally require extremely stable sample properties or dilute probes to overcome the diffraction limitation. Complex optics and sensitive instrumentation have thus far limited widespread use of microscopic particle tracking in NP gels. However, the value of visualization at the nanoscale is driving rapid advancement. Given that percolated NP gels can show extensive heterogeneity at the mesoscale, microscopy could also prove useful for resolving long-range heterogeneity and coarsening in gel networks. Moreover, tracking tracer particles in dilute NP networks may provide direct quantification of anomalous motion in complex matrices.

\subsection{Rheology}
Rheology explores the response of a material to a change in applied stress or shear-rate. Steady-state and transient configurations enable rheological experiments to probe NP gel dynamics resulting from various perturbations~\cite{petekidis2021}. Start-up shear experiments apply a constant shear rate to a material at rest and measure the material's stress response. Step-stress applies a sudden change in shear stress and measures the material's creep compliance. Oscillatory shear simultaneously probes the material's elastic storage and loss moduli as a function of angular frequency. 

Rheological measurements probe an impressively wide range of time scales, making them a powerful tool for characterizing the dynamic mechanical response of NP gels. While commercial oscillatory shear rheometers typically operate in the range of 0.1--100 Hz, some instruments can extend measured time scales to multiple days~\cite{virag2024}. On the opposite end, specialized high-frequency instruments can extend the frequency range into the ultrasound regime ($\leq$20 kHz)~\cite{schroyen2020}.

The rheological response of colloidal gels has been theoretically linked, through the mechanical relaxation spectrum $H(\tau)$, to the colloid cluster-size distributions predicted to be realized due to different types of interactions and gelation pathways~\cite{zaccone2014, song2023}. Unfortunately, directly extracting $H(\tau)$ from rheological measurements remains extremely challenging~\cite{baumgaertel1992interrelation,stadler2009new,bae2015logarithmic}, as is reconstructing colloid cluster-size distributions from experimental scattering data~\cite{liu2011lysozyme,godfrin2013intermediate,jadrich2015origin,bollinger2016fluids,godfrin2018dynamic}. As a result, understanding how microscopic interactions, structure, and dynamics relate to the mechanical relaxation spectrum remains a major challenge. An extensive discussion of some of the modeling efforts that have been developed to interpret experimental rheological data for gels and other soft materials can be found elsewhere~\cite{song2023,petekidis2021}.

DLS measurements of colloidal dispersions can be compared to shear rheology to understand how aggregation or microscopic dynamics influence viscoelastic properties~\cite{gomez-merino2014, kohl2020, dasilva2011}. Moreover, in situ DLS measured during material shear deformation has been used to more directly relate dynamic relaxation to macroscopic mechanical response~\cite{pommella2019, pujala2018}.

Similarly, in-plane microscopy carried out during rheological experiments has been used to directly observe macroscopic phase transitions or fluid deformations in response to shear~\cite{hartge2024,zhao2022,dong2022}. Pre-built, rheo-microscopy instruments have been commercially developed to enable real-space visualization of structural rearrangements while measuring mechanical response~\cite{zotero-item-4766}. In situ rheo-optical measurements have also driven the development of custom rheology cells, with recent techniques enabling simultaneous microscopy and DLS measurements during shear between parallel transparent plates~\cite{chiara2025}. While optical limitations prevent observation of nanoscopic responses to bulk shear, the application of DDM in these setups could be used to analyze how slow, long-range network relaxations are influenced by macroscopic perturbations.

In situ rheo-SAXS is a widely used method for linking bulk mechanical response to underlying structural perturbations. For example, rheo-SAXS has been employed to help explain the shear-thinning behavior of fluid dispersions of clay platelets~\cite{hotton2024, philippe2011} and cellulose nanorods~\cite{munier2022, ghanbari2025} by observing collective reorientation with the flow field. Simultaneous measurement of rheology and small-angle neutron scattering has similarly provided insights into the microstructural origins of the mechanical properties of aging silica NP gels~\cite{gordon2017rheology}. Rheo-SAXS has recently become more accessible, with pre-built benchtop configurations now commercially available.~\cite{zotero-item-4764}.

Rheology and XPCS have been used in parallel across multiple systems~\cite{guo2010connecting,guo2011,bahadur2019,liu2021,chen2021}, demonstrating powerful complementarity for understanding soft NP materials.
In situ rheology during XPCS measurements (rheo-XPCS, as shown in Figure~\ref{AI_rheoxpcs}a) is a promising new technique used to reveal how nanoscopic dynamics contribute to macroscopic network mechanics. Rheo-XPCS at the Advanced Photon Source in Argonne National lab is already proving its versatility in measuring different soft materials. Chen et al. have used rheo-XPCS to investigate the shear-thinning behavior of nanoplatelet dispersions~\cite{chen2020}, emergence of elasticity and nonergodicity in dynamically heterogeneous nanoclay glasses~\cite{chen2020}, and aging-induced mechanical hysteresis in thermoreversible colloidal silica gels~\cite{chen2023}. Rheo-XPCS has also been used to microscopically characterize yielding behavior~\cite{he2025}, mechanical response ~\cite{horwath2024, he2024}, and microscopic origins of structural and rheological memory~\cite{kamani2025} in model dispersions of colloidal silica.

Rheo-XPCS is an important advancement in the experimental techniques available at coherent light sources. In situ measurements of dynamics across length scales during mechanical shear are the most direct method of connecting NP gel dynamics to their elasticity, yielding behavior, and frequency-dependent mechanical response. Fourth-generation light sources further extend these techniques to the measurement of fast, multiscale nonequilibrium dynamics during bulk material shear. Application of rheo-XPCS to NP gel networks could provide valuable insight into how complex heterogeneous dynamics arising from multi-scale heterogeneity, reconfigurable bonding, and gel aging translate from the nanoscale to macroscopic network mechanical properties.

\subsection{Simulation methods across scales}
To simulate the large length and time scales relevant to NP gelation, coarse-grained representations of the solvent and NP surfaces are required. Implicit solvent methods, particularly Brownian dynamics (BD), are widely used because of their simple implementations and computational speed~\cite{ermak_brownian_1978, doyle_brownian_2005}. It is possible to simulate up to $O(10^6)$ NPs with BD~\cite{fenton2023minimal}. 

The most common implementation of BD assumes that the hydrodynamic drag force on each particle is equal to that of an isolated particle (i.e., $-6\pi a \mu U$; the Stokes drag force on a spherical particle of radius $a$ moving at velocity $U$ in a Newtonian solvent of shear viscosity $\mu$), which neglects all hydrodynamic interactions (HI) among particles. 
Thermodynamic behavior does not depend on HI, and accurate predictions of thermodynamic properties from BD (or any other simulation approach) only require an adequate model for conservative interactions and sufficient configurational sampling.
However, because HI significantly modify NP trajectories, so-called ``freely-draining" (FD) implementations of BD that neglect HI cannot accurately describe many dynamic quantities when compared to experimental observations. BD-FD simulations are typically inaccurate in their predictions of dominant dynamic assembly and relaxation pathways and their time scales, the structure of metastable gels that depend strongly on the kinetics of their formation, and rheology, sedimentation coefficients, and other transport properties~\cite{varga_hydrodynamic_2016, varga_hydrodynamics_2019, torre_hydrodynamic_2023}. Dynamic predictions from algorithms that include HI are significantly more accurate. Various approaches have been developed for BD simulations that include many-bodied far-field HI that decay slowly with interparticle distance $r$ (e.g., $1/r$ for the slowest decaying mode)~\cite{fiore_rapid_2017, fiore_rapid_2018, rotne_variational_1969}, pairwise near-field lubrication interactions~\cite{silbert_colloidal_1997, silbert_rheology_1999, li_simulating_2024, kumar_origins_2010}, and both far-field and near-field hydrodynamics (``Stokesian dynamics'')~\cite{brady_stokesian_1988, sierou_accelerated_2001, fiore_fast_2019, torre_python-jax-based_2025}. In mesoscale solvent methods, solvent molecules are represented by fewer and simpler effective particles that are intermediate in size between solvent molecules and dispersed NPs.  The effective particles exchange momentum with submerged species, propagate solvent flow, and mediate HI. Dissipative particle dynamics (DPD)~\cite{groot_dissipative_1997, espanol_perspective_2017} is the most common mesoscale solvent method for simulating NP gel networks.  Because of the increased computational time required to simulate mesoscale solvent relative to free-draining approaches, the maximum number of NPs that can currently be modeled by algorithms like DPD is $O(10^4)$~\cite{nabizadeh_life_2021}.  Simulations of this size may capture some, but certainly not all, of the structural and dynamic heterogeneity in gels. BD-HI and DPD are considered the state-of-the-art methods for colloidal gels. Most simulation studies in the literature have focused on spherical NPs, but both BD-HI and DPD techniques have rigid composite-body implementations for modeling other NP shapes\cite{delmotte_modeling_2025}. Another methodology, multiparticle collision dynamics, has not yet been applied to NP gels, but it is poised to be comparable in accuracy and speed to BD-HI and DPD as the technique continues to be developed~\cite{howard_modeling_2019}. Thorough discussion of these and other simulation techniques can be found in method-focused reviews~\cite{bolintineanu_particle_2014, maxey_simulation_2017, wagner_methods_2021}.

Simulations are useful for modeling the early stages of gelation by computing structural order parameters (e.g. radial distribution functions, structure factors, light scattering intensities, cluster size distributions, or contact/coordination number distributions) and material properties (e.g. shear moduli or optical extinction spectra) over time as an initially disordered suspension assembles into a gel network. To ensure accuracy and numerical stability, each time step of a simulation must resolve the fastest dynamic mode, which is typically the diffusion of a single NP over a length scale comparable to its size. For a $\SI{100}{nm}$ diameter NP, this characteristic diffusion time $\tau_D$ is approximately $\sim\SI{1}{ms}$. The longest time scale accessible to the simulation is limited by the allocated computational time, and simulations typically extend to $10^5 \tau_D$. Aging of gels occurs on time scales many orders of magnitude longer than computationally accessible time scales; e.g., $1$ day of aging is about $O(10^8 \tau_D)$ for a $\SI{100}{nm}$ diameter NP. Simulating the long-term aging of NP networks remains an open challenge.

For NP simulations large enough to adequately model gel structure, evaluating material properties can present a severe computational bottleneck. For example, computing the ionic conductivity~\cite{kadulkar_transport_2020} or the extinction spectrum~\cite{sherman2024} each require a separate simulation per configuration. Given the interest in NP-based optical metamaterials, substantial recent effort has gone into developing electromagnetic simulations to model light-matter interactions of NP assemblies~\cite{sherman_illuminating_2024}.  Moment-based methods, particularly the coupled dipole method, are the fastest approaches and readily model gel configurations containing $O(10^5)$ spherical NPs. Boundary element methods are also quite efficient and can simulate up to $O(10^3)$ arbitrarily-shaped NPs (e.g. nanorods), but have not yet been applied to NP gels~\cite{solis_toward_2014}. BD and DPD simulations of NP gel assembly typically use triply-periodic boundary conditions and allow for some degree of overlap of NP surfaces, so care must be taken for electromagnetic simulations to properly handle these aspects. Recently, an implementation of the coupled dipole method, the mutual polarization method, which accounts for triply-periodic configurations with NP overlaps, has been used to accurately model the far-field and near-field optical properties of NP gels self-assembled with BD simulations. These studies predicted how the extinction spectrum evolves during gelation~\cite{sherman2023}, how NP size, composition, and concentration change spectral properties~\cite{kang2023, kang2023a}, and how light absorption is distributed heterogeneously throughout the gel microstructure~\cite{sherman2024}. 

Beyond the slowest evolving mode characterizing the overall gelation kinetics, a variety of intrinsic time scales can be probed in simulation via time correlation functions, including mean-squared displacement, dynamic structure factor, intermediate and self-intermediate scattering functions, and other quantities described in Section III. For some correlation functions, a gel's ensemble-average response can be decomposed into contributions from individual NPs. For example, van Hove self-correlations, the probability $P(\Delta r, \Delta t)$ of observing a particle displacement of size $\Delta r$ over a lag time of $\Delta t$, are straightforward to compute in simulation and reveal dynamic heterogeneity in NP gels -- a subset of particles diffuse and rearrange quickly while another subset moves slowly~\cite{puertas_dynamical_2004, darjuzon_brownian_2003, nabizadeh_network_2024, mokshin_dynamic_2011}.
Slow dynamics are associated with particles localized within glassy and nearly-rigid clusters, which must move collectively. Much effort has been made to identify these clusters based on microstructure alone, which is much easier to do for simulated trajectories with perfectly defined $xyz$ coordinates compared to microscopy images~\cite{whitaker2019, nabizadeh_network_2024}.

Beyond cluster analysis, collective motion has also been investigated in simulations by computing the normal modes (or equivalently, phonon modes or vibrational density of states) of the gel network. Though, in principle, normal modes can be extracted directly from a high-resolution time series of the fluctuations in NP positions, it is common to use an approximate method that requires only a static configuration. This approximate method replaces true NP interactions with effective harmonic springs and then solves an eigenvalue problem for the collection of particle displacements from a static configuration that are eigenvalues of the equations of motion associated with, e.g., BD-FD or BD-HI~\cite{varga_normal_2018}. Provided one can extract the $xyz$ coordinates of each NP, this method can be applied to both simulated and experimental configurations. This approach has been used on confocal microscopy data for gels of micron-sized colloidal particles~\cite{rocklin2021}, though confocal imaging cannot be used for gels of nanoscale particles $10$-$\SI{100}{nm}$. Normal mode analysis reveals that colloidal gels contain many large wavelength, long time scale floppy modes~\cite{varga_normal_2018, rocklin2021}. HI contribute to the preponderance of large wavelength modes, as far-away network chains interact with one another via long-ranged HI, mediated by the intervening solvent. Simulation studies directly comparing the distribution of normal modes with and without HI reveal that models neglecting hydrodynamics overpredict the prevalence of fast, small wavelength modes and underpredict the prevalence of slow, large wavelength modes~\cite{varga_normal_2018, rovigatti_vibrational_2011, mizuno_structural_2021, mizuno_phonon_2022}. Including HI results in a relatively flat distribution of short- and long-wavelength fluctuations. Flat normal mode distributions are suggested in some experimental measurements~\cite{rocklin2021, lohr_vibrational_2014, gratale_vibrational_2016}, though the dynamic range in these experiments is small. These features are consistent with the observation that many particles are found in locally glassy clusters that do not have fast relaxation modes~\cite{whitaker2019}. The superposition of these normal modes, each of which has a single time scale, is often cited as a source of heterogeneity that leads to, for example, the stretched exponential decay Eq.~\eqref{kww} of the intermediate scattering function and other time correlation functions~\cite{cho2020, krall1998}. However, a direct method to compute time-correlation functions or rheological properties from the distribution of normal modes has not been established.

Rheological simulations also probe gel dynamics by deforming the network structure over a range of time scales and measuring the stress response, or vice versa, imposing a time-dependent stress and measuring the deformation. A straightforward approach is to perform small-angle oscillatory shear (SAOS) simulations across a range of frequencies $\omega$, thereby probing network dynamics on time scales of $1/\omega$. The dynamic range of time scales accessible to BD and DPD simulations is quite broad and is mainly restricted on the low frequency (long time) side by the total length of the simulation~\cite{johnson_influence_2019}. One-simulation-per-frequency is the typical strategy and can be trivially parallelized across many compute nodes. Alternatively, ``chirp'' methods use a single simulation with an oscillation frequency that varies over time in a manner that allows the entire viscoelastic spectrum to be computed at once~\cite{wang_surface_2019, ghiringhelli_optimal_2012, curtis_validation_2015}. A simple way to directly relate the linear viscoelasticity obtained from SAOS measurements to a time scale in the gel network is to identify the crossover frequency at which the viscous and elastic responses cross one another in magnitude or change their scaling with respect to $\omega$. Such transition frequencies can indicate, e.g., the dynamic time scale below which elastic bonds break and the network is dissipative or the time scale below which components do not have enough time to respond to deformations and therefore behave elastically\cite{bouzid_computing_2018, park_structural_2015}. However, linear viscoelasticity is not representative of the full rheological response of most NP gels. Because the NP microstructure changes significantly with deformation, gels typically have large nonlinear rheological responses, and they deform plastically above a yield stress or yield strain. Additionally, gel aging and other internal dynamical processes result in ``thixotropy'' --- rheological responses that are time-dependent. Nonlinear rheological behavior of NP gels and their dynamics on large length scales can be probed with simulations of medium amplitude and large amplitude oscillatory shear, MAOS and LAOS~\cite{moghimi_colloidal_2017, park_structural_2015, park_rheological_2020}. Other time-dependent rheological tests, like the start-up of shear~\cite{boromand_structural_2017} or creep in response to a suddenly applied stress~\cite{johnson_yield_2018}, have also been simulated. An important benefit to probing rheology computationally is that the signals can be directly related to the microstructure. As described in the previous section, this is typically accomplished experimentally using rheo-SAXS or related setups, in which a beam passes through a rheometer sample to obtain simultaneous measurements of rheology and scattering. With simulations, additional structural order parameters and features can be computed during rheological deformations, including the distributions of contact/coordination number~\cite{johnson_influence_2019}, cluster statistics~\cite{nabizadeh_network_2024}, and alignment/orientation measures~\cite{varga_large_2018} to learn structure-property relationships. Care must be taken to incorporate periodic boundary conditions during deformation (e.g., using Lees-Edwards boundary conditions) and to properly incorporate imposed flow into the equations of motion~\cite{allen_computer_2017, lees_computer_1972}. Additionally, while the thermodynamic contribution to rheological calculations is present in the virial there are additional stress contributions from particles and the fluid that must be accounted for via hydrodynamic calculations; see, e.g.~\cite{wang_surface_2019, wagner_methods_2021}.

While time scales associated with particle motion have been the focus of most computational investigations, time scales associated with other components, e.g., the ligands on the surface of NPs, are also important. For micron-sized particles, these molecular time scales are well-separated from the time scales for particle motion. However, for NPs $10$--$\SI{100}{nm}$, the time scales can become comparable. In simulation investigations, the time correlation of interparticle ``bonds'' has been a useful quantity to characterize short length scale rearrangements~\cite{del_gado_network_2008, del_gado_microscopic_2010}. When molecular details of particle surfaces are neglected (as is common in simulations of micron-sized particles), two particles are defined as ``bonded'' if they are within some cutoff distance typically defined from the shape of their mutual interparticle potential. The bond ``lifetime'' is the characteristic time scale over which the bond between two NPs persists. This provides some insight into how fast the overall network structure can coarsen and age (e.g., the gel cannot age faster than the rate at which bonds break and rearrange). It also relates to the network's rheological properties. Small deformations occurring faster than the bond lifetime drive an elastic response, whereas particles can rearrange and dissipate deformations slower than the bond lifetime. Defining bonded pairs through the proximity of particle centers is useful for highly coarse-grained models of particles, but for nanoscale particles where molecular details of the surface are important, bonding should be defined in terms of the proximity and chemical state of the surface groups. For example, several NP gelation strategies involve dynamic covalent or metal-ligand coordination bonding between surface groups. Simulations of linker-mediated NP gelation have quantified how molecular bonding time scales are distinct from, but intimately coupled with, particle time scales~\cite{kwon_dynamics_2022}. These results provide insights for experimental investigations in which well-controlled dynamic covalent bonding time scales govern the rheological response of NPs~\cite {dominguez2020, kang2022} and polymer networks~\cite {fitzsimons2022, crowell2023}.

Though the dynamics of surface groups are important to model as NP size decreases, explicitly including NPs, surface ligands, ions, and solvent in a single simulation is not generally feasible. Even coarse-grained implicit-solvent models using Kuhn-sized bead/spring representations of surface ligands are restricted to $O(100)$--$O(1000)$ NPs over time scales not much larger than the NP diffusion time~\cite{dominguez2020,kang2022}. Additionally, incorporating chemical reaction kinetics, which may drive gelation via bonding of surface groups between two NPs, into dynamic simulations is challenging because such reactions can occur on time scales that may differ substantially from those of NP motion~\cite{holoman_simulating_2025}.  These aspects make it difficult to directly relate the large-scale, long-time features of NP gels (which determine rheological properties and aging) to the molecular species that drive gelation. An open problem in molecular modeling is to resolve the multiscale nature of NP gelation. Similarly, the ultra-long-time behavior of NP gels on scales of days, weeks, and even months is inaccessible to computer simulations designed to resolve single-NP dynamics ($\SI{1}{ms}$ for $\SI{100}{nm}$ particles). It would be extremely useful if simulation models could provide accelerated aging predictions, as experiments at these time scales are necessarily slow.

For highly coarse-grained models, $O(10^6)$ particles are possible to model, making it feasible to access length scales much larger than the particle size. However, even this is not necessarily adequate to describe the fractal nature of gels. A major issue is that particle gels are only fractal over intermediate length scales much larger than the finite size of individual particles and much smaller than the finite size of the material sample. The common approach of using the box-counting dimension to determine a network's fractal dimension is sensitive to system size and error-prone, especially because small changes in fractal dimension can lead to pronounced changes in material properties. Reaching sufficiently large length scales is especially challenging for networks of nonspherical NPs. Faceted, rod-like, and other anisotropic NPs are typically modeled via composite rigid bodies, so a single NP can require $\ge O(100)$ discrete elements to accurately model.\cite{delmotte_modeling_2025} This challenge is amplified for rheological simulations, which (on top of the large length scales required) also require large time scales to impose deformations. Researchers are improving the computational efficiency of composite body simulations, and methods for simulating true-faceted particles that do not require composite formulations are promising~\cite{shi_energy-conserving_2025}.

\subsection{Experimental considerations and the self-consistency problem}
Any single experimental technique is insufficient to fully characterize multi-scale heterogeneous gel dynamics. XPCS is rapidly becoming the standard for studying the evolution of microstructure and emergent dynamics in NP gels, but the increasing capabilities of various techniques to resolve dynamic phenomena across space and time are promising for developing a more complete picture of NP gel dynamics and for streamlining gel investigations prior to beamline experiments. In any event, it is crucial to respect the limitations of each technique when attempting a coordinated characterization of NP gels.

First, multiple scattering restricts DLS to the characterization of dilute NP dispersions. Cross-correlation in 3D-DLS extends measurable volume fractions, but high local density in NP gels may distort measured dynamics if care is not taken to tailor the scattering path length to the sample opacity. Despite this limitation, DLS fits within a broader experimental framework by enabling rapid access to the effects of primary particle design on diffusive processes. Measuring NPs with varied size, shape, core properties, and surface modifications in different solvents with DLS provides a means to investigate and compare their respective influences on NP dynamics. Similarly, 3D-DLS can be used to probe how long-range thermodynamic and hydrodynamic interactions modulate particle motion across length scales in concentrated dispersions, bridging dilute suspensions and arrested gel networks.

Microscopy is sensitive to optical noise, particularly in scattering or opaque samples. The current microscopy capabilities are consequently better suited for studying NP interactions outside of fully formed gels, where they can provide real-space insight into how clusters form or rearrange. Microscopy may also help evaluate how mesoscale structure and heterogeneous coarsening emerge during assembly. The appeal of direct nanoscopic visualization is likely to continue to drive advances in particle-tracking and imaging modalities, with the long-term benefit of enabling reliable measurements in higher-density NP dispersions or arrested gel networks.

Since DDM uses microscopy to measure statistics of intensity-fluctuation correlations, it can substantially exceed the optical limitations of direct image analysis. DDM has not yet been widely used to study smaller NPs, but its flexible optical conformation, contrast mechanisms, and analysis methods leave substantial room for adaptation to NP gels. DDM also provides direct network visualization via real-space microscopy. With continued development, DDM could serve as a valuable extension of techniques for dynamical characterization of slow microscopic relaxations in NP gels.

Even with appropriate application of the discussed techniques, NP gels are sensitive to environmental conditions. Caution should be used when comparing dynamics measured across different samples. The intuitive remedy is to combine multiple techniques for in situ characterization. This is a broad experimental challenge; for example, integrating optical techniques requires precise optical alignment and cross-compatibility among detectors, light sources, and sample optical properties. 

Extracting accurate properties from simulations to interpret and design experiments requires addressing the challenge of model validation. Unfortunately, the structural and dynamic comparisons required to validate models are complicated by limitations from both experiments and simulations. Gels simulated in finite, periodic simulation cells cannot fully reproduce the large-scale heterogeneity of macroscopic experimental samples, and simulations cannot yet reach the long time scales available to experiments. Furthermore, while structural evolution can be characterized using small-angle X-ray speckle patterns, scattering-vector limits and ensemble averaging obscure details needed to fully describe network heterogeneity. Significant advances in simulation capabilities and in direct experimental visualization of NP network assembly, alongside the increasingly powerful scattering techniques being deployed, seem necessary to address the shortcomings of current modeling strategies.

In addition to developing and validating robust simulation models, methodical characterization of experimental dynamic properties relies on self-consistent analysis across measurements. The development of tunable model gels, real-space NP gel network visualization, and in situ techniques that provide complementary spatiotemporal information will be critical to advancing our understanding of emergent properties. 

\section{Outlook: Leveraging modular design within a systematic experimental framework}

\subsection{Ligand-linker gels as model systems}

\begin{figure}[h!]
    \centering
    \includegraphics[width = 0.5\textwidth]{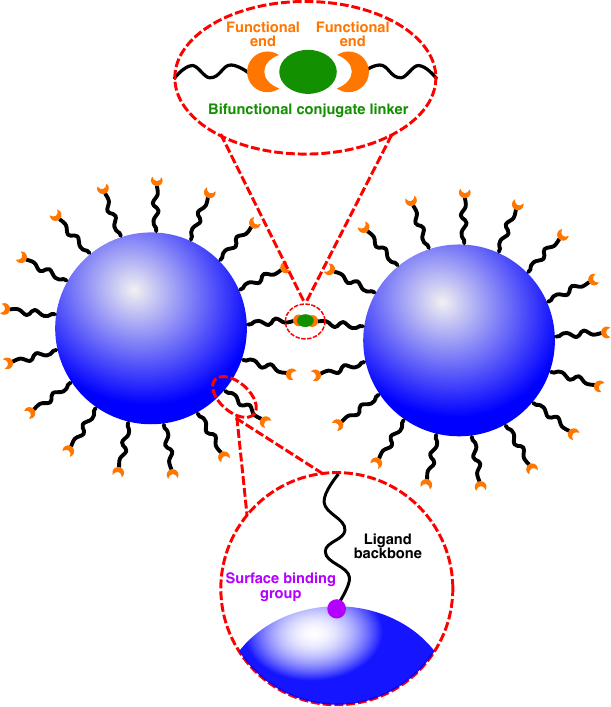}
    \caption{One linkage in a model linker gel consisting of inorganic NP cores (blue) with ligands composed of variable binding groups (purple), ligand backbones (black), ligand functional ends (orange) and bifunctional linkers (green). \label{fig:linkergel}}

\end{figure}

Gelation mediated via dynamic bonding between ligand-functionalized NPs and bifunctional small-molecule linkers (Fig.~\ref{fig:linkergel}) is a modular and tunable strategy for assembling NP gels\cite{green2022, sherman2021}. ``Linker gels'' are powerful model systems because it is possible to independently change properties of NPs (size, shape, and composition), ligands (molecular weight, NP surface coverage), and linkers (molecular weight and concentration) for a given ligand-linker bonding chemistry. This control has been leveraged to isolate the effects of many of these physical features on NP gel morphology and material properties. Similarly, there is a growing pool of ligand-linker chemistries that can be selected independent of NP and ligand physical features, providing a handle to tune chemical bonding equilibrium, kinetics, and lifetime~\cite{green2022}.

A group of four chemical reactions has been identified to be mutually tunable, orthogonal, reversible, and covalent (TORC). Bis-terpyridine/metal complexes, aldehyde/hydrazide, boronic esters, and thia-Michael linkages are TORC chemistries that are active under mild conditions and have been used to form highly controllable polymer networks~\cite{reuther2019}. Bifunctional linker/ligand pairs designed with  these TORC chemistries form the basis for promising model NP gels. The equilibrium and kinetic properties of TORC bonding can be systematically changed by altering nucleophilic or electrophilic character of reactants, enabling control over bond energetics~\cite{Seifert2016}. TORC bonds have facilitated rational control of polymer network mechanics by modulation of dynamic bond exchange rates and equilibria~\cite{fitzsimons2022,richardson2019,terriac2024,peng2020}. 

TORC bonding allows explicit and orthogonal control over NP attractions while preserving core and ligand structure. Bond enthalpy and exchange kinetics can be systematically tuned by modifying the reversible bond chemistry. Linker concentration or competitive “capping" species can be used to shift bond equilibrium and clustering behavior~\cite{dominguez2020}. End-functionalized linear or star-polymer linkers provide additional design handles to enhance interparticle bond elasticity. Asymmetric linkers with different TORC functionalities on each end could be used to control mixed NP network composition. Dynamic covalent chemistry represents a rich landscape for programmable NP gel design, offering pathways to probe gelation physics and engineer microscopic dynamics, optical, and network mechanical properties~\cite{singh2022}.

TORC chemistry also enables variation of gelation mechanism across similar NPs. Thermoreversible terpyridine-metal-linked indium tin oxide (ITO) nanocrystal (NC) gels have demonstrated explicit tunability of gelation temperature by shifting terpyridine-metal coordination equilibria~\cite{kang2022}, selecting different metal and halide combinations to modify competitive terpyridine-metal and halide-metal equilibria~\cite{kang2025}, or modifying NP ligand length~\cite{kang2023}. Such systems are strong candidates for parameterized characterization of thermoreversible NP gels with controlled properties. XPCS has not yet been applied to these systems; slow thermal quenches or ultrafast XPCS at XFEL facilities could potentially observe the how anomalous dynamics emerge during gelation~\cite{guo2011,bahadur2019}.

Temporally evolving gels linked via dynamic chemistries may provide platforms for real-time investigation of emergent heterogeneous or nonequilibrium structure and dynamics. Using aldehyde-functional ligands and bifunctional hydrazide linkers, ITO NC gels have been formed on time scales ranging from hours~\cite{ofosu2025} to weeks~\cite{dominguez2020}, demonstrating widely tunable gelation kinetics. Thus far, gelation kinetics have shown dependencies on the concentrations of linker and an added non-coordinating salt.

XPCS and SAXS measurements throughout gelation to study the latter mechanism of acceleration have revealed universal evolution of structure and dynamics across different salt-controlled time scales~\cite{ofosu2025}. This work sets a precedent for high-time-resolution XPCS experiments to evaluate how nanoscopic parameters or additives affect dynamic properties in gels approaching arrest and during subsequent network aging. Furthermore, upgraded X-ray sources and complementary techniques offer a route to fundamentally resolve gelation pathways and achieve a comprehensive characterization of how NP design governs gelation properties across scales.

In summary, linker-gels can be used as a platform for controlled experimentation of isolated parameters within four main categories:
\begin{itemize}
    \item{Primary particle effects: NP size, shape, and composition}
    \item {NP surface modification effects: particle connectivity, binding group charge, and ligand hydrodynamics}
    \item {Linker-ligand pairing: chemical bond equilibria and kinetics, linker architecture, and symmetry of linker functionalization}
    \item {Additional handles for tunability: salt, depletants, and capping species}
\end{itemize}

\subsection{Spectroscopic tracking of structural evolution in plasmonic NP gels}
For metal or doped metal oxide NPs, collective near-field coupling of localized surface plasmon resonances upon gelation modulates optical extinction in a manner that reflects underlying structure~\cite{kang2023,kang2025, dominguez2020, ofosu2025, sherman2021, saezcabezas2018,sherman2023}. Comparing the optical properties and structure factors of simulated and experimentally prepared gels provides a means to validate models and elucidate gel formation pathways.

The localized surface plasmon resonance of a small, quasispherical, plasmonic nanocrystal features a single near-symmetric peak at the resonant frequency~\cite{agrawal2018}. During gelation, the peak shifts to lower frequencies and becomes asymmetrically broadened, reflecting enhanced coupling to other NPs. 
Tracking changes in the spectral lineshape enables characterization of structural evolution and gel-formation kinetics. In a recent study of linked ITO nanocrystal gels, time-dependent monitoring of optical extinction was combined with information about microstructural evolution measured by SAXS during XPCS. This analysis revealed universal salt acceleration of gel formation via a time-salt superposition~\cite{ofosu2025}.

Simulating dynamic changes of the optical response of model plasmonic NPs upon gel formation is possible using the mutual polarization method, as discussed in Section IV ~\cite{sherman2023}. The relative simplicity and accessibility of UV-Vis-NIR spectroscopy make this approach attractive for learning about network coarsening and emergent structural heterogeneity inaccessible to scattering measurements. In situ UV-Vis-NIR measurement of gels undergoing shear may allow direct correlation of dynamic mechanical response to emergent network-scale structure~\cite{singh2022}. 

\subsection{Ultra-resolution nanoscopy}

While predominantly used for imaging biological networks and cellular environments, super-resolution microscopy (SRM) has been conservatively applied to NP systems. Structured illumination microscopy has been used to resolve plasmonic gold NP assemblies on 2D substrates~\cite{wu2024}, and stochastic optical reconstruction microscopy has probed interactions between proteins and silica NPs~\cite{feiner-gracia2017}. Fluorescent quantum dots~\cite{jin2018} and plasmonic NPs~\cite{li2022} have also been proposed as tracer particles in biological media and highlight the potential of SRM for NP systems. However, systematic application of SRM to inorganic NP gels remains limited due to challenges in labeling, photo stability, and image contrast imposed by the intrinsic properties of inorganic NPs.

Historically, electron microscopy of liquids has been limited to samples confined between thin electron-transparent membranes. Advances in electron microscopy optics, membrane material, and cell construction continue to improve the compatibility of ultra-resolution electron microscopy with soft colloidal materials~\cite{Bharda2019}.

Liquid-phase transmission electron microscopy (LPTEM) could become a powerful method to probe gel dynamics and directly image microscopic gel structures. Unlike conventional TEM, LPTEM images NPs in their native liquid environment confined within a sealed microfluidic cell. In just the past five years, LPTEM has been used to directly visualize sterically trapped gold and platinum NPs~\cite{lajer2025}, probe transport and interaction dynamics of gold NPs in graphene liquid cells~\cite{Kang2021}, and capture various metal NP nucleation and growth phenomena~\cite{kim2025,zhang2024}. Although LPTEM has not yet been extensively employed to study NP gels, the technique represents a promising front for real-time visualization of gel assembly, cluster formation, and network rearrangements at nanometer resolution~\cite{shabeeb2025}.

Electron-beam-induced artifacts remain a central challenge to accurately interpreting intrinsic gel dynamics. Careful dose-minimization strategies and optimized detector settings are critical for preserving the dynamic properties and structure of NP gels. Inorganic NPs are generally resilient to electron beam damage, but organic components, including ligands, solvents, and molecular linkers, are more vulnerable. Temporal resolution in LPTEM is similarly constrained by dose limitations and camera sensitivity, imposing trade-offs between imaging speed, observation time, and sample integrity.

While individual NP dynamics in dilute systems can be analyzed via simple single-particle tracking, resolving dynamics in dense colloidal networks remains challenging due to image congestion and particle overlap. Manual particle identification is infeasible, and simple algorithmic tracking fails to capture motion in crowded environments. Recent advances in machine-learning-based image analysis now enable extraction of stochastic dynamic information from LPTEM videos~\cite{shabeeb2025}. Improvements in LPTEM cell engineering and image analysis provide local probes to help understand the origins of the dynamics observed in XPCS.

\subsection{Increasing efficiency of high-throughput and nonequilibrium data analysis during XPCS}

XPCS has become the centerpiece for characterizing the dynamics of nanostructured soft materials. Brookhaven National Laboratory, Argonne National Laboratory, the European Synchrotron Radiation Facility, and PETRA III (DESY) now provide experimental space dedicated to XPCS. Additionally, X-ray free electron lasers (XFELs) such as the European XFEL, Linac Coherent Light Source, and Spring-8 Angstrom Compact Free-Electron Laser generate highly coherent X-rays, enabling ultrafast and ultrahigh-resolution scattering measurements.

While high resolution detectors with fast frame rates are necessary for accessing fast dynamics in NP gels, continuously measuring an evolving gel generates massive volumes of data. A study at the Advanced Photon Source (APS) before the fourth-generation upgrade records the data flux of XPCS with a 20 $\mu$s resolution and a 500k pixel Rigaku detector at 0.2 petabytes per 24 hours of operation~\cite{zhang2021}. Cloud integration of XPCS data workflows with supercomputing capabilities addresses strenuous data processing to provide real-time feedback for experimenters~\cite{vescovi2022, chard2025}. 

Streamlined user experiences at XPCS facilities offer flexibility when measuring many samples. Conveniences like Python-based control, intuitive user interfaces~\cite{chu2022}, multi-sample stages, and automated robotic sampling~\cite{ozgulbas2023} continue to maximize efficient use of light sources and can support thorough parameterization of NP gel dynamics.

Despite these advances, extensive human intervention is still required to interpret nonequilibrium XPCS data. Bottlenecks for quantitative analysis of spatially and temporally heterogeneous NP systems are exacerbated by the high volumes of data generated during XPCS experiments. Rheo-XPCS poses particular challenges due to driven nonequilibrium conditions and the need to extract dynamics without conventional time-averaging to avoid smearing fundamental dynamic processes. Two complementary methodologies have recently emerged to address these limitations: AI-driven analysis and analytical frameworks for interpreting the TTC.

\begin{figure}
    \centering
    \includegraphics[width = 0.5\textwidth ]{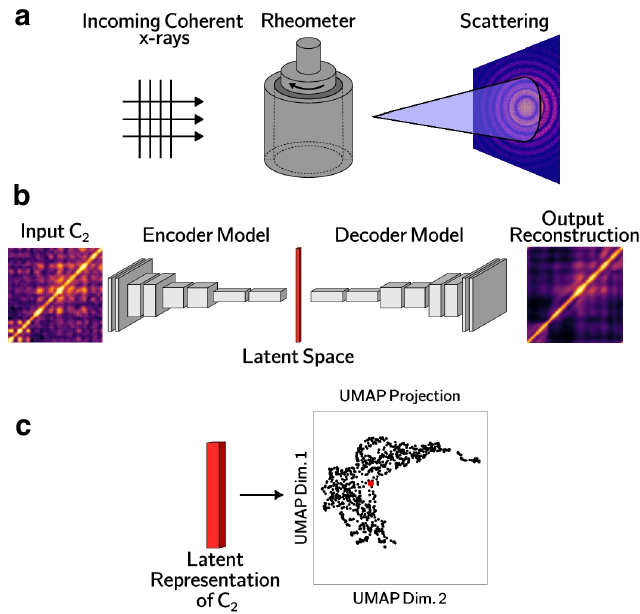}
    \caption{(a) Schematic of in situ rheo-XPCS. (b) TTC workflow using AI encoder models to analyze nonequilibrium TTC function. (c) Latent space classification of nonequilibrium on a two-dimensional Uniform Manifold Approximation and Projection (UMAP) diagram. Adapted from~\cite{horwath2024}, licensed under CC BY 4.0.}
    \label{AI_rheoxpcs}
\end{figure}

Horwath et al.~\cite{horwath2024} constructed a training dataset from the scattering patterns of spherical silica NPs in polyethylene glycol across varying volume fractions and shearing conditions. A convolutional autoencoder projects TTCs into a latent space, enabling automated classification of dynamics in sheared colloidal glasses based on similarity to learned patterns (Figure~\ref{AI_rheoxpcs}b). This categorization automates identification of distinct dynamical regimes and facilitates association of dynamic signatures with variables such as volume fraction, particle size, shear stress, and measurement position (Figure~\ref{AI_rheoxpcs}c).  AI-driven XPCS analysis promises significant reduction of manual TTC interpretation and could accelerate high throughput experimental series to discover the origins of nonequilibrium dynamic behavior in NP gels and other soft matter.

He et al.~\cite{he2024} developed an analytical framework for nonequilibrium rheo-XPCS that extracts dynamic information without averaging across sample age. Their approach relates TTCs to a characteristic transport coefficient using a Langevin-based description of both thermal and externally driven motion. After validation on simulated thermally quenched non-interacting colloids, the method is applied to experimental systems. They successfully quantify transport coefficients for rheo-XPCS data from silica suspensions~\cite{donley2023}, elastically linked iron-oxide NP assemblies under mechanical strain~\cite{song2022}, and ligand-functionalized silica particles exhibiting shear-banding transitions. In each case, the framework enables quantitative interpretation of TTCs and direct extraction of physically meaningful dynamic parameters within their model.

Emerging AI and analytical methods provide the means to interpret complex XPCS datasets and connect nanoscale dynamic process to bulk mechanical response. As XPCS capabilities grow, such approaches will be essential for understanding multiscale dynamics in NP gels and establishing mechanistic links between evolving network properties and their nanoscopic origins.

\section{Conclusion}
Colloidal NP gels exhibit complex dynamic behavior that is not captured by a single relaxation mechanism or structural scale. Their evolution is instead governed by coupled hierarchical processes spanning NP nanoscopic motion, mesoscopic cluster rearrangements, and collective network relaxations. The intrinsic multi-scale nature, many-body interactions, mesoscopic heterogeneity, and nonequilibrium relaxation challenge comparisons between experimental NP gels and simulated models of colloidal dynamics. At the heart of the matter, dynamic processes in NP gels span spatiotemporal regimes that far exceed the reach of any single technique.

In this perspective, we discuss current interpretations of anomalous and nonequilibrium dynamics investigated with scattering-based methods. Fourth-generation light sources and advanced detectors now allow XPCS to use this analysis to gain insight into fundamental processes ranging from nanoscopic motion to microscopic relaxation. Light scattering, microscopy, simulations, and rheology provide a framework for complementary analysis of NP gel dynamics across a broader range of spatiotemporal scales. Improved experimental workflows and data analysis at coherent light sources promise accelerated systematic investigation of NP gel dynamics, while new rheo-XPCS offers a route to elucidate the relationships between nonequilibrium mechanical response and its nanoscopic origins. By applying this framework to NP gels with precise tunability, we can begin to answer fundamental physical questions about how complex multi-scale dynamics govern gelation and emergent network properties.

A systematic approach to investigating inorganic NP gels is required to establish design principles and enable bottom-up construction of tunable or adaptive materials. By linking anomalous, heterogeneous, and nonequilibrium gel dynamics to their nanoscopic origins, these phenomena are reframed as physically meaningful features that can be exploited to control emergent network properties. More broadly, these insights contribute to the development of general and physically grounded models across diverse NP systems.

\section*{Acknowledgment}
This work was primarily supported by the National Science Foundation (NSF) through the Center for Dynamics and Control of Materials: an NSF MRSEC under Cooperative Agreement No.~DMR-2308817 and under an NSF DMREF project, CBET-2323482. F.L. acknowledges funding by the Deutsche Forschungsgemeinschaft (DFG, German Research Foundation) as part of the Excellence Strategy of the Federal Government and the federal states - EXC 3120/1 BlueMat: Water-Driven Materials – project number 533771286.

\bibliography{ref}
\end{document}